\documentclass[%
 reprint,
superscriptaddress,
 amsmath,amssymb,
 aps,
]{revtex4-2}

\usepackage{graphicx}
\usepackage{dcolumn}
\usepackage{bm}
\usepackage{multirow}
\usepackage[colorlinks,plainpages=false,linkcolor=blue,urlcolor=blue,citecolor=blue,pdfpagemode=UseNone]{hyperref}
\usepackage[dvipsnames]{xcolor}
\usepackage{ulem}

\newcommand{\cofetof}{CoFe$_{2}$O$_{4}$}
\newcommand{\xcfo}{(Co$_{1-x}$Fe$_{x}$)$^{\rm{Tet}}$(Co$_{x}$Fe$_{2-x}$)$^{\rm{Oct}}$O$_{4}$}
\newcommand{\mub}{$\mu_{\rm B}$}

\newcommand{\gzwz}{$G_{0}W_{0}$}
\newcommand{\etal}{{\sl et al.}}

\newcommand{\wrt}{with respect to}

\begin{document}

\preprint{APS/123-QED}

\title{Electronic and optical properties of the fully and partially inverse CoFe$_{2}$O$_{4}$ spinel from first principles calculations including many-body effects}

\author{Shohreh Rafiezadeh}
\affiliation{Department of Physics and Center for Nanointegration Duisburg-Essen (CENIDE), University of Duisburg-Essen, Duisburg, Germany}

\author{Vijaya Begum-Hudde}
\affiliation{Department of Materials Science and Engineering, University of Illinois, Urbana-Champaign, Urbana, IL 61801, USA}
\affiliation{Department of Physics and Center for Nanointegration Duisburg-Essen (CENIDE), University of Duisburg-Essen, Duisburg, Germany}

\author{Rossitza Pentcheva}
\affiliation{Department of Physics and Center for Nanointegration Duisburg-Essen (CENIDE), University of Duisburg-Essen, Duisburg, Germany}

\date{}

\begin{abstract} 
Using density functional theory (DFT) calculations and state-of-the-art 
many-body perturbation theory, we investigate
the electronic and optical properties of the inverse spinel CoFe$_{2}$O$_{4}$, a common anode material for photocatalytic water splitting. 
Starting with different exchange-correlation functionals, at the independent particle level we obtain a direct band gap of 1.38~eV (PBE+$U$, $U$~=~4~eV)
and 1.69 eV (SCAN+$U$, $U$ = 3~eV), whereas HSE06 renders an indirect band gap of 2.02~eV. 
Including quasiparticle effects within $G_{0}W_{0}$, a larger and indirect band gap is obtained for all functionals: 1.78~eV (PBE+$U$, $U$=~4~eV),  1.95~eV (SCAN+$U$, $U$ = 3~eV) and 2.17~eV (HSE06) which is 29\%, 15\% and 5\% higher than the independent particle (IP) band gap, respectively. Excitonic effects, taken into account by solving the Bethe-Salpeter equation (BSE) lead to a redshift of the optical band gap to 1.50 (SCAN+$U$, $U$ = 3~eV) and 1.61~eV (HSE06), in good agreement with the reported experimental values  1.50$-$2.0~eV. 
The lowest optical transitions in the visible range, identified by means of oscillator strength, are at 2.0, 3.5, and 5.0~eV, consistent with experimental observations at 2.0, 3.4, and 4.9~eV.
We also explored the effect of the degree of inversion:  the band gap is found to decrease from 1.69 ($x=1$) to 1.45 ($x=0.5$), and 1.19~eV ($x=0)$  within the IP approximation with SCAN+$U$, $U$=3~eV. This trend is reversed after the inclusion of excitonic effects, resulting in a band gap of 1.50, 1.57, and 1.64~eV  for $x$ = 1.0, 0.5, and 0.0, respectively.
The oscillator strength analysis of the BSE calculations indicates that both $x$ = 0.0 and $x$ = 0.5 exhibit transitions below 1~eV with extremely small oscillator strengths that are absent in the inverse spinel. This corroborates previous suggestions that these transitions are due to the presence of Co$^{2+}$ cations at the tetrahedral sites.
\end{abstract}
\maketitle


\section{\label{sec:Introduction}Introduction}
The high demand for large-scale green energy production 
has led to an increased necessity for photocatalysts with optimized performance~\cite{dresselhaus2001alternative,chu2012opportunities}.
Iron and cobalt oxides such as Fe$_{2}$O$_{3}$, Co$_{3}$O$_{4}$, NiFe$_{2}$O$_{4}$ and \cofetof (CFO) are promising 
candidates due to  their desirable properties, such as abundance, low cost, high chemical stability under reaction conditions and optical transitions in the visible range
~\cite{henrich1996surface,hajiyani2018surface,peng2021influence,PhysRevLett.103.176102,chakrapani2017role}.
Therefore, a fundamental knowledge of their optical properties is essential for the catalyst selection process and opens up possibilities for further optical applications.

\begin{figure*}[!htp]
\includegraphics[width=.95\textwidth]{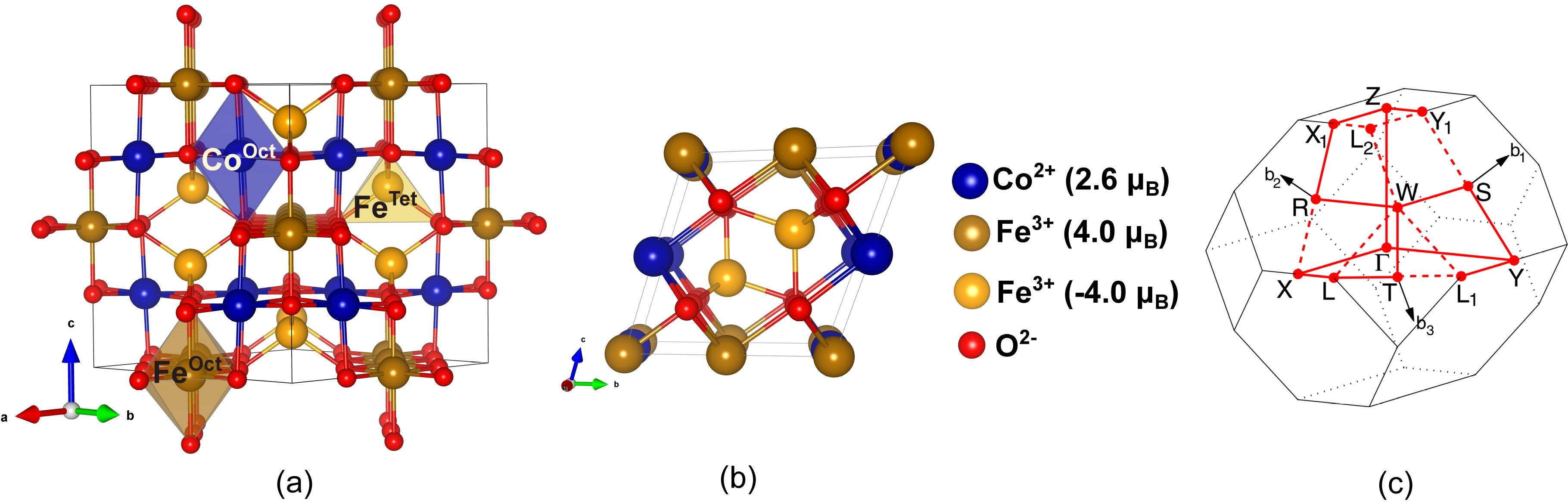}%
\caption{\label{fig:structure} Crystal structure of the inverse spinel CoFe$_{2}$O$_{4}$: 
(a) conventional unit cell, containing 56 atoms, with octahedral (Co$^{\rm{Oct}}$ and Fe$^{\rm{Oct}}$)
and tetrahedral (Fe$^{\rm{Tet}}$) sites
(b) primitive unit cell with two formula units, 14 atoms, 
and (c) high symmetry points in the Brillouin zone of CoFe$_{2}$O$_{4}$, adopted from AFLOW \cite{Curtarolo2012-AFLOW}.}
\end{figure*}

In this work, we investigate the electronic and optical properties of the inverse spinel
cobalt ferrite CoFe$_2$O$_4$.  
In this compound equal amounts of Fe$^{3+}$ ions occupy octahedral (Oct) and tetrahedral (Tet) sites, with antiparallel orientation of the spins on the two sublattices. The ferrimagnetic nature of this material with Curie temperature $T_{\rm C} = 790$~K~\cite{schmitz2013electric} stems from the ferromagnetically ordered Co$^{2+}$ ions which are located at octahedral sites, as shown in Fig.~\ref{fig:structure}. However, in real samples some degree of disorder can occur which is described by the degree of inversion $x$, quantifying the fraction of divalent Co cations occupying octahedral sites. The chemical formula, describing the partially inverted spinel, is  \xcfo{} where $x$~=~1 presents the perfect inverse spinel structure. The magnitude of $x$ can significantly influence the electronic, magnetic, and optical properties of spinels~\cite{Venturini2019-INV,Granone2018-INV-znfe2o4,Sharma2022-INV}. %

Despite numerous experimental and theoretical studies~\cite{schmitz2013electric,zheng2004multiferroic,zavaliche2007electrically,hajiyani2018surface,kampermann2021link}, the optical properties of CFO, as well as the effect of the degree of inversion, are still debated. For example, 
 Holinsworth \etal{}~\cite{Holinsworth2013-opt-gap} obtained a direct gap of 2.80~eV at 4.2K and 2.67~eV at 800K using the Tauc plot approach.
 By using the same method Himcinschi~\etal{}~\cite{Himcinschi2013-DF} reported a direct gap of 1.95~eV, whereas, Kalam~\etal\ \cite{Kalam2018-opt-gap} and Ravindra~\etal\ \cite{Ravindra2012-opt-gap} obtained an optical band gap of 2.5-2.65~eV.
In addition, Dileep \etal{}~\cite{Dileep2014-bstr-mBJLDA} reported a direct optical gap of 2.31~eV by employing spatially resolved high resolution electron energy loss spectroscopy. 
Singh \etal{}~\cite{Singh2018-opt-gap} measured an optical gap of 1.65~eV.
which reduces to 1.55~eV and 1.43~eV upon applying a magnetic field of 400 and 600~Oe, respectively.
Sharma and Khare~\cite{Sharma2014-opt-gap} reported an optical gap of 1.58~eV ($T$ = 500$^{\circ}$) and 1.41~eV ($T$ = 700$^{\circ}$) for \cofetof{} films deposited on quartz substrates. 
Recently, Singh~\etal{}~\cite{Singh2020-opt-gap} showed that the optical band gap decreases from 1.9 eV to 1.7 eV with increasing nanoparticle size. The larger nanoparticles were found to have a cation distribution similar to bulk CFO with respect to the degree of inversion.
Overall, the reported experimental direct optical gaps show a wide variation between $\sim$0.55 - 4.1~eV. Moreover, optical transitions below 1~eV have been related to crystal field transitions of tetrahedrally coordinated Co$^{2+}$ which is not present in the fully inverse spinel \cite{Fontijn1999-off-diagonal-DF}.

Besides the variation in the measured band gaps, theoretical calculations also show a wide range of values from 0.52  to 1.90~eV~\cite{Fritsch2010-DFT,Dileep2014-bstr-mBJLDA,Holinsworth2013-opt-gap,Lukashev2013-DFT-gap,Dimitrakis2016-dft-gap}, depending on the method and exchange-correlation functional used.
Density functional theory (DFT) calculations within the generalized-gradient approximation (GGA) fail to render the insulating state and predict a half-metallic behavior instead~\cite{Paji2004-dft-gap}. Dileep~\etal{}~\cite{Dileep2014-bstr-mBJLDA} calculated a total indirect band gap of 0.80 eV in the minority channel using the modified Becke-Johnson local density approximation. Using GGA in the parameterization of Perdew, Burke, and Enzerhof (PBE) ~\cite{Perdew1996-PBE} with an on-site Hubbard Coulomb term $U_{\rm{Co}}$=$U_{\rm{Fe}}$= 3~eV on the Co and Fe $3d$ electrons 
an indirect band gap in the minority spin channel of 0.52~eV was obtained at the GGA lattice parameter\cite{caffrey2013spin} using the VASP code~\cite{kresse199614251,kresse1999ultrasoft} and 0.80~eV~\cite{Lukashev2013-DFT-gap} with the QUANTUM
ESPRESSO (QE) code~\cite{Giannozzi2009-QE}. For the GGA+$U$ lattice parameter, a larger band gap of 0.90~eV~\cite{Fritsch2010-DFT} was reported.  On the other hand,  a direct gap of 1.08 eV between the minority valence band and majority conduction band was found by Pemmaraju~\etal{}~\cite{pemmaraju2007atomic}, employing an atomic self-interaction correction (ASIC) scheme.
In contrast, an indirect band gap of 1.60 eV~\cite{caffrey2013spin} was obtained with the hybrid functional(HSE03)~\cite{Heyd2003-HSE}. While most theoretical studies have focused on the inverse spinel, 
Sharma~\etal{}\cite{Sharma2022-INV} reported recently that the band gap of 1.09~eV (PBEsol+$U$,$U_{\rm{Co}}$=$U_{\rm{Fe}}$=~4~eV) for the inverse spinel is reduced by 6\%  for $x$ = 0.5. Much smaller band gaps, decreasing from 0.72~eV ($x$ = 1.0) to 0.1 ($x$ = 0.0), were found by Hou~\etal{}~\cite{Hou2010-inversionT} using PBE+$U_{\rm{eff}}$ ($U_{\rm{Co}}$= 4.08, $J_{\rm{Co}}$ = 0.79~eV and $U_{\rm{Fe}}$= 4.22, $J_{\rm{Fe}}$ = 0.80~eV).

An improved description of the electronic structure beyond DFT can be achieved by considering many-body effects, e.g. by calculating the quasiparticle energies by means of the self-energy as a product of the single-particle Green’s function $G$ and the screened 
Coulomb interaction $W$,
in the $GW$ approximation introduced by Hedin~\cite{hedin1965new}. 
The single shot \gzwz{} was shown to yield a good description of the band gap of other spinels such as Co$_3$O$_4$~\cite{Smart2019-co3o4-polarons,singh2015putting} and ZnFe$_2$O$_4$~\cite{Ulpe2020-BSE-inversion}.
An important aspect is the starting point of the $GW$ calculation. In particular for transition metal oxides as well as rare earth compounds, adding an on-site Coulomb term within   LDA(GGA)+$U$  renders a better description than LDA or GGA ~\cite{Jiang2009-G0W0-LDAU,Jiang2010-GW-LDAU,jiang2012ab,lany2013band,piccinin2019band,Ulpe2020-BSE-inversion}.
For example, Lany \cite{lany2013band} showed that employing a Hubbard $U$ term significantly improves the $GW$ band structure for a series of
nonmagnetic, antiferromagnetic and ferrimagnetic transition metal
compounds.

Electron-hole interactions can significantly influence the optical spectrum. These can be taken into account by solving the Bethe-Salpeter equation (BSE)~\cite{Rohlfing1998-BSE,Rohlfing1998-BSE}. This generally improves the agreement with experiment regarding the spectral features and energetic positions of the peaks in a wide range of (transition) metal oxides such as  ZnFe$_{2}$O$_{4}$~\cite{Ulpe2020-BSE-inversion}, MgAl$_{2}$O$_{4}$~\cite{jiang2012ab}, LiCoO$_{2}$~\cite{Radha2021-LiCoO2}, $\alpha$-Fe$_{2}$O$_{3}$~\cite{piccinin2019band}, SrTiO$_{3}$~\cite{Sponza-2013,Begum2019-p1} and MgO~\cite{Wang2004,Begum2021-p2}.

To our knowledge, \gzwz\ and BSE have not been applied previously to CFO. In this work, starting from different exchange correlation functionals, we have calculated the optical spectrum of CFO including quasiparticle corrections within the single-shot \gzwz\ and excitonic corrections by solving BSE. To evaluate the effect of different exchange-correlation functionals  
on the electronic and optical properties of bulk CFO, we have employed the GGA (PBE) and the  strongly constrained and approximately normed meta-GGA (SCAN)~\cite{Sun2015-SCAN} 
functionals with different Hubbard $+U$ values,
as well as the hybrid functional HSE06~\cite{Krukau2006-HSE06}. 
To gain a deeper understanding of the impact of cation distribution and in particular the degree of inversion on the optical properties of  CoFe$_2$O$_4$,
we calculated the optical spectra additionally for $x$ = 0.5 and 0.0.
Due to the high computational demand of \gzwz+BSE calculations, we also tested a model BSE scheme (mBSE) with lower computational cost for the treatment of static screening~\cite{Bokdam2016-MBSE,Fuchs2008-mBSE,Liu2018-mBSE,Bechstedt1992-mBSE}. This approach has been applied  previously to SrIrO$_{3}$ (also Sr$_{2}$IrO$_{4}$ and Sr$_{3}$Ir$_{2}$O$_{7}$)~\cite{Liu2018-mBSE}, SrTiO$_{3}$~\cite{Begum2019-p1}, MgO~\cite{Begum2021-p2} and to a set of transition metal oxide perovskites like SrTiO$_{3}$, SrMnO$_{3}$ and LaVO$_{3}$~\cite{Varrassi2021-mBSE} with an overall good agreement between the mBSE and \gzwz+BSE spectra.

The paper is structured as follows: In Sec. \ref{sec:CompDetail} the computational details are presented. The results are discussed in Sec. \ref{sec:Result}. Specifically, Sec. \ref{subsec:properties} is dedicated to the ground state structural and electronic properties of CFO, whereas Sec. \ref{subsec:QP-bstr} presents the quasiparticle (QP) band structure. Sec. \ref{subsec:OpBstr} discusses the optical properties of CFO, in particular, 
we present and analyze the real and imaginary parts of the dielectric function and the absorption coefficient at different levels of treatment with different starting exchange-correlation functionals. 
Finally, in Sec. \ref{subsec:Inversion} we assess the effect of the degree of inversion and cation distribution on the structural, electronic, and optical properties of CFO. 
The results are summarized in Sec. \ref{sec:Summary}. In Appendix \ref{sec:mBSE}, the spectrum of the inverse spinel obtained with model BSE is compared to the \gzwz+BSE spectrum.

\begin{table*}[!htbp]
\caption{\label{tab:table1} The calculated lattice constant $a$ (\AA), magnetic moment ($\mu_{\rm{B}}$) and band gap of \cofetof{} in the independent particle (IP) and quasiparticle (QP) approximation, using PBE+$U$, SCAN+$U$, and HSE06 functionals with a  Hubbard $U$ term applied on the Co and Fe 3$d$ states. $E_{\rm{gap}}^{\rm{TOT}}$ is  
the total band gap in eV. $E_{\rm{gap}}^{\rm{dn}}$ and $E_{\rm{gap}}^{\rm{up}}$ are the band gaps in the minority and majority spin channels, respectively. Index "(i)" and "(d)" denotes an indirect and direct band gap, respectively.}
\begin{ruledtabular}
\begin{tabular}{lccccccccccc}
&\multicolumn{1}{c}{} & \multicolumn{3}{c}{magnetic moments ~($\mu_{\rm{B}}$)} & \multicolumn{3}{c}{IP} &
\multicolumn{3}{c}{$G_{0}W_{0}$}  \\ 
 Functional/$U$ in eV & $a$ (\AA)   & Co$^{\rm{Oct}}$ & Fe$^{\rm{Tet}}$ & Fe$^{\rm{Oct}}$  &   $E_{\rm{gap}}^{\rm{up}}$  & $E_{\rm{gap}}^{\rm{dn}}$ & $E_{\rm{gap}}^{\rm{TOT}}$  & $E_{\rm{gap}}^{\rm{up}}$ & $E_{\rm{gap}}^{\rm{dn}}$ & $E_{\rm{gap}}^{\rm{TOT}}$ \\ \hline
PBE+$U$, $U_{\rm{Co}}$=$U_{\rm{Fe}}$= 3  &   8.39      &  2.65    &  -4.08   &   4.21  & 1.74$_{(\rm{i})}$  &  0.92$_{(\rm{i})}$  &  0.92$_{(\rm{i})}$   &  2.54$_{(\rm{i})}$  &  1.32$_{(\rm{i})}$  &  1.32$_{(\rm{i})}$  \\
PBE+$U$, $U_{\rm{Co}}$=$U_{\rm{Fe}}$= 4 &   8.40     &  2.70    &  -4.29   &   4.41   & 2.05$_{(\rm{i})}$ & 1.53$_{(\rm{i})}$ & 1.38$_{(\rm{d})}$ & 2.68$_{(\rm{d})}$ & 1.78$_{(\rm{i})}$ & 1.78$_{(\rm{i})}$  \\
PBE+$U$, $U_{\rm{Co}}$= 5, $U_{\rm{Fe}}$= 4 &   8.39      &  2.78    &  -4.30   &   4.42 &  2.28$_{(\rm{i})}$  &   2.48$_{(\rm{i})}$   &  1.49$_{(\rm{d})}$  &  2.83$_{(\rm{d})}$   &  2.75$_{(\rm{i})}$  &  2.07$_{(\rm{d})}$  \\
SCAN+$U$, $U_{\rm{Co}}$=$U_{\rm{Fe}}$= 3  &   8.33      &  2.70    &  -4.22   &   4.32   & 2.45$_{(\rm{i})}$ & 1.72$_{(\rm{i})}$  & 1.69$_{(\rm{d})}$ &  3.08$_{(\rm{i})}$ & 1.95$_{(\rm{i})}$  & 1.95$_{(\rm{i})}$  \\
SCAN+$U$, $U_{\rm{Co}}$=$U_{\rm{Fe}}$= 4 &   8.34      &  2.77    &  -4.40   &   4.51  & 2.75$_{(\rm{i})}$ & 2.31$_{(\rm{i})}$  & 2.11$_{(\rm{d})}$ & 3.34$_{(\rm{d})}$ & 2.39$_{(\rm{i})}$ & 2.39$_{(\rm{i})}$  \\
HSE06      &   8.37     &  2.61    &  -3.98   &   4.09  & 3.30$_{(\rm{i})}$ & 2.02$_{(\rm{i})}$  & 2.02$_{(\rm{i})}$  & 3.65$_{(\rm{i})}$ &  2.17$_{(\rm{i})}$ & 2.17$_{(\rm{i})}$ \\
\multirow{1}{*}{Experiment}
&  ~~8.39\footnote{Reference~\cite{Singh2020-LC,martens1982-DF-CFO,li1991single}.}      &            &            &       &  &  &  &   &  &  &    \\
\end{tabular}
\end{ruledtabular}
\end{table*}

\section{\label{sec:CompDetail}Computational details}

The calculations were performed using the projector augmented wave (PAW) method~\cite{blochl1994projector,kresse1999ultrasoft} 
implemented in the Vienna {\sl ab initio} simulation package (VASP)~\cite{kresse199614251,kresse1999ultrasoft}, and employing PAW pseudopotentials, specially designed for $GW$ calculations. For the exchange correlation functional, we have used  PBE~\cite{Perdew1996-PBE}, SCAN~\cite{Sun2015-SCAN}, and HSE06~\cite{Krukau2006-HSE06}.
For PBE and SCAN an additional on-site Hubbard 
Coulomb repulsion parameter $U_{\rm{eff}}$= $U-J$  is applied to the Co and Fe $3d$ states
within the Dudarev \etal\ \cite{Dudarev-HubbardU} approach.
The electron configurations of Co, Fe and O are  3$d^{8}$4$s^{1}$, 3$d^{7}$4$s^{1}$ 
and 2$s^{2}$2$p^{4}$, respectively. 
The conventional cubic spinel unit cell of CFO with $Fd\bar{3}m$ space group contains eight spinel formula units.
We have used the primitive rhombohedral unit cell including two spinel formula units 
with 14 atoms (Fig.\ref{fig:structure}~b) to reduce the computational cost. The plane-wave cutoff energy is set to 500 eV.
For the integration over the Brillouin zone, we use a $\Gamma$-centered 5 $\times$ 5 $\times$ 5 $\mathbf{k}$ mesh. Both the volume and the internal parameters were optimized with the residual forces smaller than 0.001 eV/\AA$^{-1}$. 
The band structures are interpolated using the WANNIER90 code \cite{mostofi2008wannier90} along the high symmetry point path 
adopted from AFLOW \cite{Curtarolo2012-AFLOW} and FINDSYM \cite{Stokes2005-FINDSYM}.
Our calculations with and without spin-orbit coupling (SOC), 
showed that Co acquires a significant orbital moment of $0.16$\mub, but the band structure is only weakly modified (see Fig. S1 in supplementary information), therefore, we 
proceed with the results without SOC.

Single-particle excitations were described in terms of electron and hole QPs by adopting the $GW$ approximation.
Within the first-order perturbation, the  DFT wave functions $\psi_{n}$ are used and  
the QP energy of a state $n$, $\varepsilon_{n}^{QP}$ is defined as:
\begin{equation} \label{eq:QP-energies}
\small{\varepsilon_{n}^{QP} = \varepsilon_{n}^{DFT} + \langle  \psi_{n} \vert \Sigma(\varepsilon_{n}) - V_{xc} \vert \psi_{n} \rangle}
\end{equation} 
where $\varepsilon_{n}^{DFT}$ and $\varepsilon_{n}^{QP}$ are 
the DFT single-particle and quasiparticle energies, respectively. 
$\Sigma(\varepsilon_{n})$ is the self-energy operator which is obtained by both Green's function $G$ and the screened Coulomb potential $W$ calculated using DFT single-particle energies and wavefunctions.
Convergence with respect to the number of bands and the cutoff energy of the response function
in the $G_{0}W_{0}$ calculations 
were ensured by employing 792 bands and a cutoff energy of 333~eV (see Fig. S2 in the supplementary information)~\cite{Dubecky2023}.

To take into account excitonic effects, 
we solve the Bethe-Salpeter equation~\cite{Hanke1980-BSE}:

\begin{equation}\label{eq:BSE-eq}
\small{
(E_{c\mathbf{k}}^{QP}-E_{v\mathbf{k}}^{QP})A^{\lambda}_{vc\mathbf{k}}+\sigma_{v^{\prime} c^{\prime} 
\mathbf{k^{\prime}}}^{QP}\langle vc\mathbf{k} \vert K^{eh} \vert v^{\prime} c^{\prime} 
\mathbf{k^{\prime}} \rangle A^{\lambda}_{v^{\prime} c^{\prime} \mathbf{k^{\prime}}} = 
\Omega^{\lambda} A^{\lambda}_{vc\mathbf{k}}}.
\end{equation}

Within the Tamm-Dancoff approximation, 
vertical transitions from the valence to the conduction  band $(E_{c\mathbf{k}}^{QP}-E_{v\mathbf{k}}^{QP})$ are considered. $\vert v^{\prime} c^{\prime} \mathbf{k^{\prime}} \rangle$ are the corresponding electron-hole pair configurations,
$A^{\lambda}_{vc\mathbf{k}}$ are the expansion coefficients on the electron-hole basis functions, 
$\Omega^{\lambda}$ are the exciton eigenenergies, 
and $K^{eh}$ is the kernel that takes into account the electron-hole interaction.
An accurate description of the optical spectrum in $G_{0}W_{0}+$BSE, 
especially the calculation of Re[$\epsilon$($\omega$)] from the Kramers-Kronig relation requires 
a large number of $\mathbf{k}$-points and empty states. For a large system like CFO, this enhances substantially the
computational time and memory demand.
In this work, the BSE calculations were performed with 24 (28) occupied (unoccupied) bands on 
a $\Gamma$-centered 5 $\times$ 5 $\times$ 5 $\mathbf{k}$ mesh 
to evaluate the electron-hole excitation energies in the range of 0$-$6 eV (see Fig. S3 in the supplementary information).
The dielectric function is evaluated by using 100 (imaginary) frequency and imaginary time grid points. A Lorentzian broadening of 0.3 eV is applied to all the calculated optical and absorption coefficient spectra to mimic the excitation lifetime.

\begin{figure*}[tb!]
\includegraphics[width=0.9\textwidth]{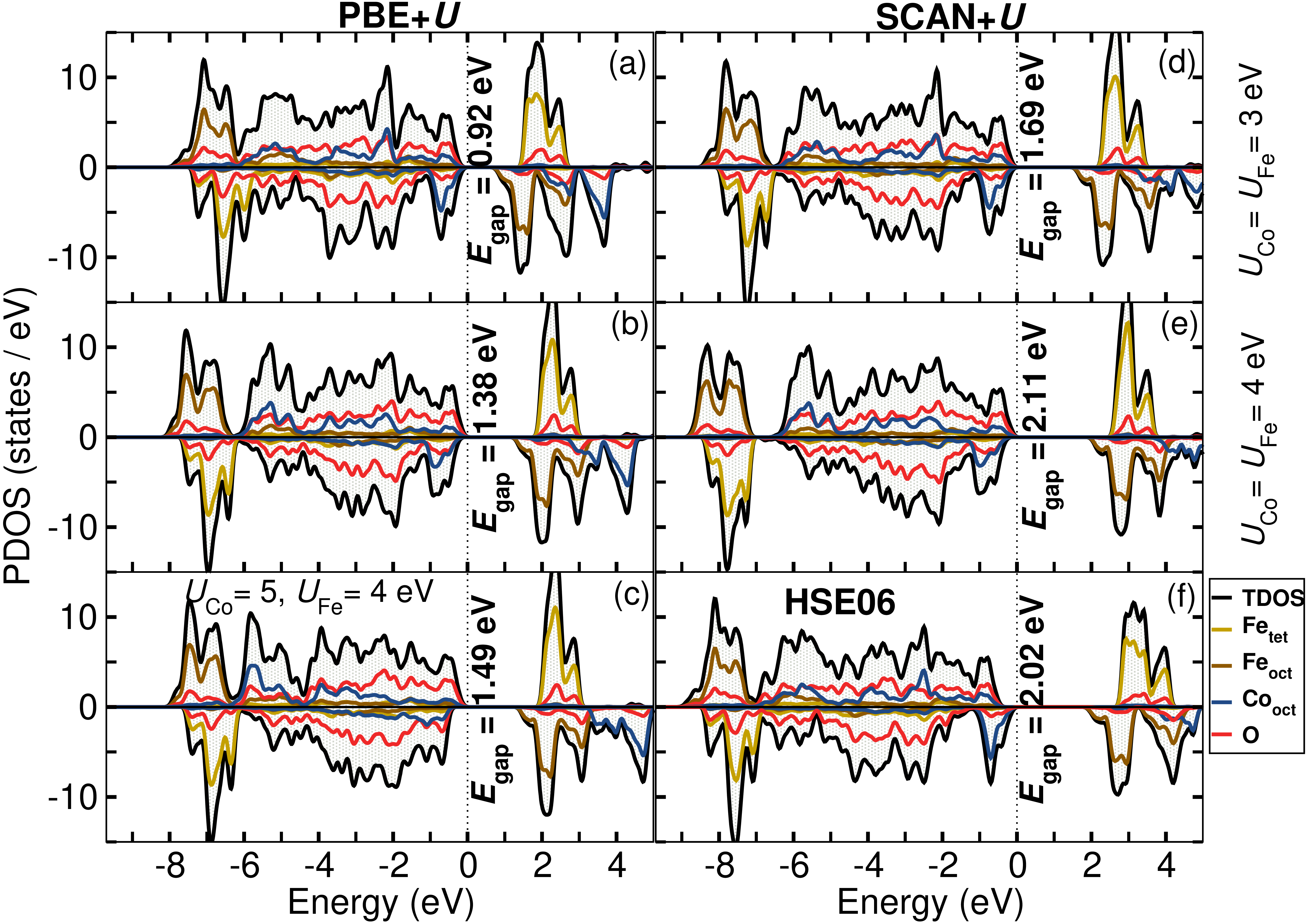}
\caption{Total and projected density of states (TDOS/PDOS) of \cofetof\ calculated with PBE+$U$: a) $U_{\rm{Co}}$=$U_{\rm{Fe}}$= 3 eV, b) $U_{\rm{Co}}$=$U_{\rm{Fe}}$= 4 eV and c) $U_{\rm{Co}}$= 5 eV, $U_{\rm{Fe}}$= 4 eV);  SCAN+$U$ d) $U_{\rm{Co}}$=$U_{\rm{Fe}}$= 3 eV, e)$U_{\rm{Co}}$=$U_{\rm{Fe}}$= 4 eV) 
and f) HSE06 functionals. Bold numbers indicate the calculated band gaps ($E_{\rm{gap}}$) in eV.
}
\label{fig:PDOS}
\end{figure*}


\section{\label{sec:Result} Results and discussion}
\subsection{\label{subsec:properties}Structural and electronic properties}
\subsubsection{Ground state structural electronic properties}

We start our analysis with the structural properties of CFO obtained with 
different starting exchange-correlation functionals, namely, PBE, SCAN including a
Hubbard $U$ term on the Co and Fe 3$d$ states and HSE06.
The lattice constants are presented in Table~\ref{tab:table1}: the PBE+$U$ value is 8.39 \AA\ ($U_{\rm{Co}}$=$U_{\rm{Fe}}$= 3 eV),
8.40 \AA\ ($U_{\rm{Co}}$=$U_{\rm{Fe}}$= 4 eV), and 8.39 \AA\ ($U_{\rm{Co}}$ = 5 eV and $U_{\rm{Fe}}$ = 4 eV) almost coinciding with 
the experimental value of 8.39 \AA~\cite{Singh2020-LC,martens1982-DF-CFO,li1991single}. 
With SCAN+$U$ the lattice parameter is 8.33 \AA\ ($U_{\rm{Co}}$=$U_{\rm{Fe}}$= 3 eV) and  8.344 \AA\ ($U_{\rm{Co}}$=$U_{\rm{Fe}}$= 4 eV), 
0.61$\%$ and 0.55$\%$ less than the experimental value, respectively.
With the hybrid functional HSE06, the 
lattice constant is 8.37 \AA\ (0.24$\%$ smaller than the experimental one).

As shown in Table~\ref{tab:table1}, all three functionals render a ferrimagnetic ground state.  The Fe$^{3+}$ at tetrahedral and octahedral sites are aligned antiparallel and cancel each other. The net magnetization of CFO stems from the ferromagnetically aligned Co$^{2+}$ at octahedral sites.
With PBE+$U$/SCAN+$U$, $U_{\rm{Co}}$=$U_{\rm{Fe}}$=~3~eV, the calculated magnetic moments 
are 2.65/2.70~$\mu_{\rm{B}}$ for Co$^{\rm{Oct}}$, 
4.21/4.32~$\mu_{\rm{B}}$ for Fe$^{\rm{Oct}}$ 
and -4.08/-4.22~$\mu_{\rm{B}}$ for Fe$^{\rm{Tet}}$. 
These values increase 
to 2.70/2.77, 4.41/4.51, and -4.29/-4.40 $\mu_{\rm{B}}$ 
for Co$^{\rm{Oct}}$, Fe$^{\rm{Oct}}$ and Fe$^{\rm{Tet}}$ for a higher $U_{\rm{Co}}$=$U_{\rm{Fe}}$=~4~eV, respectively.
With HSE06 the magnetic moments are slightly lower 2.61, 4.09, and -3.98 $\mu_{\rm{B}}$ for Co$^{\rm{Oct}}$, Fe$^{\rm{Oct}}$ and Fe$^{\rm{Tet}}$, respectively. 
The total magnetic moment of 3~$\mu_{\rm{B}}$ per formula unit with all three functionals is in good agreement with the experimental value of 3.25~$\mu_{\rm{B}}$~\cite{shahbahrami2022exploring}.

The total-, element- and orbital-projected density of states (TDOS and PDOS) obtained with the different exchange-correlation functionals is presented in Fig.~\ref{fig:PDOS}~(a-f).
While the bottom of the valence band  (-6.0 to -8 eV) is dominated by Fe~3$d$ states at the tetrahedral (minority spin channel) and octahedral sites (majority spin channel), Co~3$d$ and O~2$p$ states prevail at the top of the valence band. However, depending on the $U$ value and the starting functional, the valence band maximum (VBM) is in the minority spin channel with PBE+$U$ ($U_{\rm{Co}}$=$U_{\rm{Fe}}$= 3~eV) and HSE06, whereas it is in the majority spin channel for PBE+$U$ ($U_{\rm{Co}}$=$U_{\rm{Fe}}$= 4) and SCAN+$U$ ($U_{\rm{Co}}$=$U_{\rm{Fe}}$= 3 and 4~eV). 
The conduction band is mostly comprised of Fe~3$d$ states at the tetrahedral (majority spin channel) and octahedral sites (minority spin channel), the conduction band minimum (CBM) for all functionals being in the minority spin channel.

In general, the band gap increases with $U$ for both PBE and SCAN: 
With PBE+$U$, the band gap is 0.92 eV for $U_{\rm{Co}}$=$U_{\rm{Fe}}$= 3 eV, 1.38 eV for $U_{\rm{Co}}$=$U_{\rm{Fe}}$= 4 eV and 1.49 eV  for  $U_{\rm{Co}}$ to 5 eV and $U_{\rm{Fe}}$= 4 eV.
SCAN+$U$ renders the same trend but significantly larger values of 1.69 and 2.11 eV for $U_{\rm{Co}}$=$U_{\rm{Fe}}$= 3 and 4 eV, respectively.
For the hybrid functional HSE06, a band gap of 2.02 eV is obtained. Further insight into the nature of calculated band gaps, as well as the position of VBM and CBM, is provided by analyzing the band structures in the following section.

\begin{figure*}[htb!]
\includegraphics[width=0.95\textwidth]{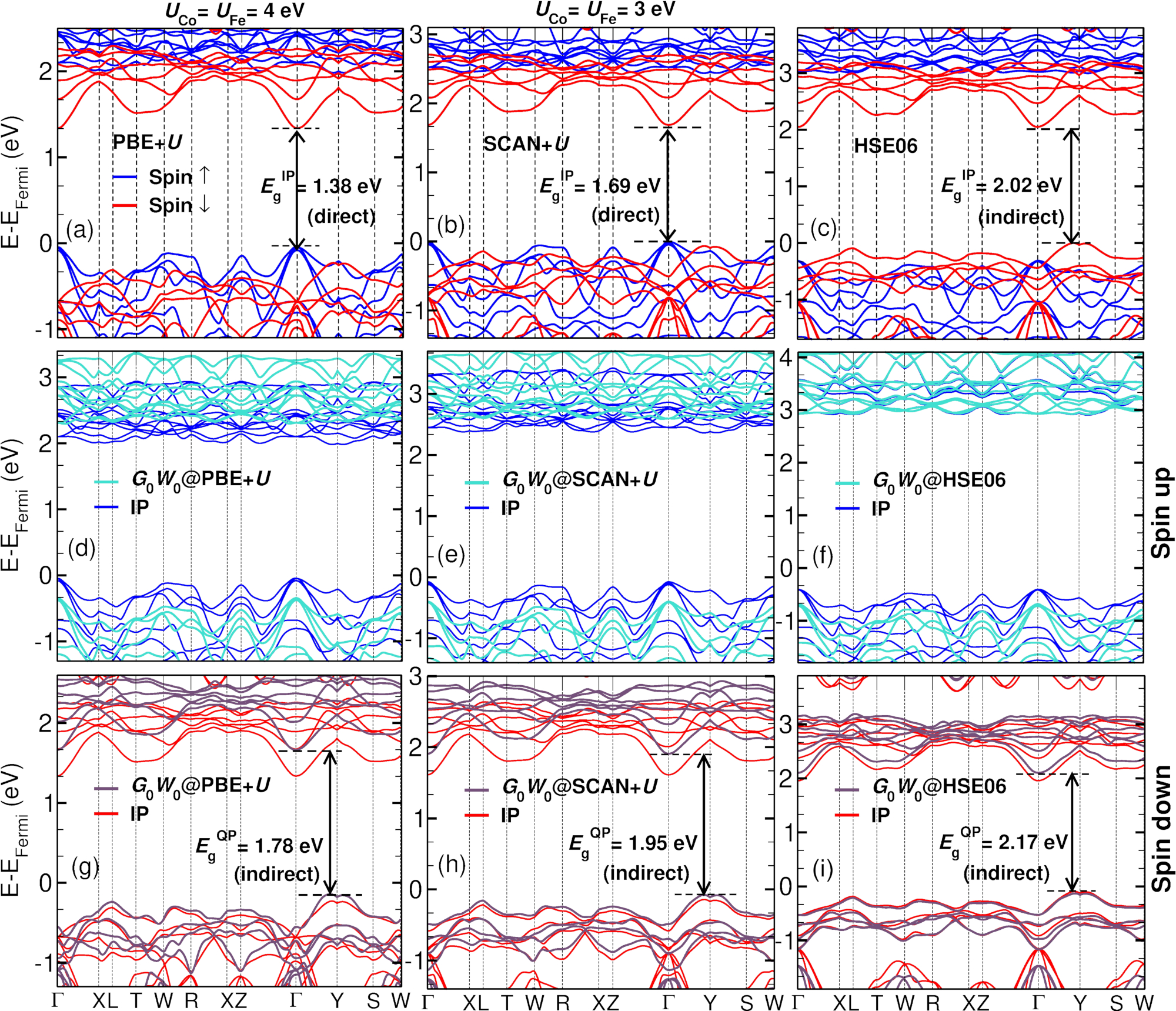}
\caption{Band structure of \cofetof{} calculated within (a, b, c) the independent particle (IP) approximation with the different functionals, PBE+$U$ ($U_{\rm{Co}}$=$U_{\rm{Fe}}= 4$ eV), SCAN+$U$ ($U_{\rm{Co}}$=$U_{\rm{Fe}}= 3$~eV) and HSE06, respectively,  (blue and red denote majority and minority spin channels), and the $G_0W_0$ approximation 
for (d, e, f) the majority and (g, h, i) minority spin channels, respectively. Maroon and cyan denote the quasiparticle (QP) band structure plotted together with the IP band structure.
}
\label{fig:bstr}
\end{figure*}
%

\subsubsection{\label{subsec:QP-bstr} Independent Particle and quasiparticle band structure}

In this section, we discuss the band structures obtained with PBE$+U$, SCAN$+U$, and HSE06 functionals within the independent particle (IP) picture and by including quasiparticle effects within $G_0W_0$, shown in Fig. \ref{fig:bstr} 
(See Fig. S5 in supplementary information for other $U$ values).
The IP band structures displayed  in Fig.~\ref{fig:bstr}~(a-c) and Fig. S5~(in supplementary information) show that both with PBE+$U$ ($U_{\rm{Co}}= U_{\rm{Fe}}$= 4 eV) 
and SCAN+$U$ ($U_{\rm{Co}}= U_{\rm{Fe}}$= 3 and  4 eV) the VBM and CBM are located at the $\Gamma$ point in the minority and majority spin channel, respectively. In contrast, with HSE06 and PBE+$U$ ($U_{\rm{Co}}= U_{\rm{Fe}}$= 3~eV) both the VBM and CBM are in the minority spin channel, the former is located along $\Gamma$-$Y$ and the latter at $\Gamma$.

Regarding the IP band gap, with PBE+$U$, we obtain an indirect gap of 0.92 eV ($U_{\rm{Co}}= U_{\rm{Fe}}$= 3 eV) in good agreement with previous reported value of 0.95 eV using the same $U$ values~\cite{Fritsch2010-DFT,Lukashev2013-DFT-gap}.
However, for higher $U$ values $U_{\rm{Co}}= U_{\rm{Fe}}$= 4~eV the band gap switches to a direct and larger one (1.38~eV). Similarly, the band gap calculated with SCAN+$U$ ($U_{\rm{Co}}= U_{\rm{Fe}}$= 3 and 4~eV) is direct and significantly higher, 1.69, and 2.11~eV, respectively, as presented in Table.~\ref{tab:table1}. An indirect band gap of 2.02~eV is obtained with HSE06.

Including QP corrections substantially modifies the band structure for the semi-local functionals. 
The valence band in the minority spin channel shifts only by 0.06-0.08~eV, but shows modifications beyond a rigid band shift, for example, change in the position and in the order and dispersion of bands at $\sim -1$~eV, at $\Gamma$ and $W$, and along $\Gamma$-$Y$ and $R$-$X$ [cf. Fig.~\ref{fig:bstr} (g, h, i)].  In the majority spin channel, the valence band moves downwards by 0.3$-$0.4~eV. As a consequence, for both PBE+$U$ and SCAN+$U$ the VBM switches from the majority to the minority spin channel upon inclusion of QP corrections. 
On the other hand, the conduction band shifts by 0.9 - 1.1~eV upwards. This leads to an overall increase of the band gap to 1.32~eV (PBE+$U$ ($U_{\rm{Co}}= U_{\rm{Fe}}=3$ eV), 1.78~eV  (PBE+$U$, $U_{\rm{Co}}= U_{\rm{Fe}}= 4$~eV), 1.95~eV (SCAN+$U$, $U_{\rm{Co}}= U_{\rm{Fe}}= 3$~eV), and 2.39~eV (SCAN+$U$, $U_{\rm{Co}}= U_{\rm{Fe}}= 4 $~eV) and a change to an indirect band gap.

In the case of HSE06 (cf. Fig.~\ref{fig:bstr}~f and i), the  QP corrections are much smaller, the VBM/CBM shift only slightly to lower/higher energy by 0.03/0.09 eV 
in the minority spin channel. In the majority channel, the CBM is largely unchanged, 
whereas the VBM is shifted downwards by 0.28 eV at $\Gamma$. Overall the band gap of 2.02 eV (HSE06) is 
enhanced by only  5\% to 2.17~eV (HSE06+$G_{0}W_{0}$). Moreover, unlike the semi-local functionals, 
the HSE06 band gap is an indirect one in the minority spin channel with the VBM at $\Gamma$-$Y$  and the CBM at $\Gamma$ [see Fig.~\ref{fig:bstr} (c, i)] at both the IP and QP levels. 
Overall, the HSE06 functional provides an improved description of the ground-state properties and the $GW$ corrections are smaller compared to the semi-local exchange-correlation functionals.


\begin{figure*}[!htbp]
\includegraphics[width=1.0\textwidth]{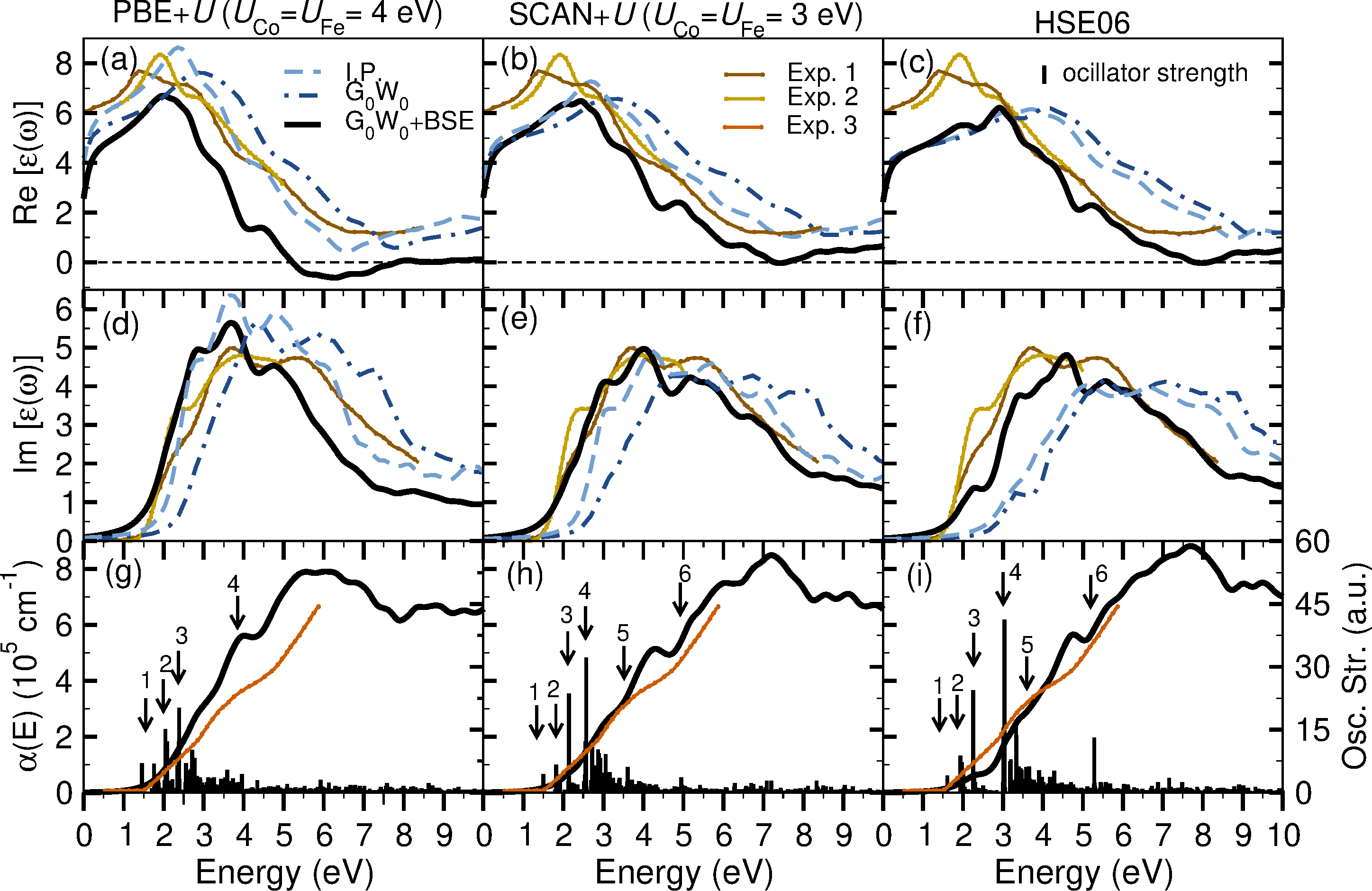}
\caption{\label{fig:opt-abs-spectra} Optical absorption spectrum of CoFe$_{2}$O$_{4}$: 
(a, b, c) real (Re[$\epsilon$($\omega$)]) and  (d, e, f) imaginary part (Im[$\epsilon$($\omega$)]) of the dielectric function 
and (g, h, i) absorption coefficient ($\alpha(E)$)
for PBE+$U$ ($U_{\rm{Co}}$=$U_{\rm{Fe}}= 4 $ eV), SCAN+$U$ ($U_{\rm{Co}}$=$U_{\rm{Fe}}= 3$ eV) and HSE06 as the starting functional, respectively, 
within IP approximation, $G_{0}W_{0}$ and $G_{0}W_{0}+$BSE. Vertical black lines denote the oscillator strength for the 
$G_{0}W_{0}+$BSE spectra.
A Lorentzian broadening of 0.3 eV is employed for all the calculated spectra. 
Experimental data are adopted from Exp. 1 \cite{Zviagin2016-DF-both}, 
Exp. 2 \cite{Himcinschi2013-DF} and Exp. 3 \cite{Holinsworth2013-opt-gap}.}
\end{figure*}

\subsection{\label{subsec:OpBstr}Optical properties}

We now turn to the optical properties of the inverse spinel CFO and discuss in detail the real, Re[$\epsilon$($\omega$)], and imaginary part, Im[$\epsilon$($\omega$)], 
of the frequency-dependent dielectric function (DF), as well as the absorption coefficient calculated 
with different starting 
exchange-correlation functionals, namely, PBE+$U$, SCAN+$U$, and HSE06 evaluated at the independent particle (IP) level and by including quasiparticle ($G_{0}W_{0}$) and excitonic effects ($G_{0}W_{0}+$BSE). The theoretical spectra are compared to the experimental results of Himcinschi~\etal\ ~\cite{Himcinschi2013-DF} and Zviagin~\etal\ ~\cite{Zviagin2016-DF-both}. These measurements were performed using ellipsometry on epitaxial films of CFO grown by pulsed-laser deposition (PLD). 
The measured real part of the DF (Re[$\epsilon$($\omega$)]) shows a peak and a shoulder at 1.38 and 2.55~eV (Zviagin~\etal~\cite{Zviagin2016-DF-both}), and at 1.95 and 2.90 eV~(Himcinschi et al.~\cite{Himcinschi2013-DF}), respectively. The experimental macroscopic static electronic dielectric constant, $\epsilon_{\infty}$ = Re [$\epsilon$($\omega$ = 0)] is 6 ~\cite{Zviagin2016-DF-both}. The measured Im[$\epsilon$($\omega$)] spectra display an onset at around 1.5~eV, a shoulder at 2.0~eV, and two broad peaks with nearly equal intensity at around 3.5 and 5~eV, and a drop in the intensity at 6.0~eV. The differences between the two studies may be related to the different substrates used: in the first case, CFO was deposited on a SrTiO$_{3}$(100) substrate at 575$^{\circ}$C\cite{Himcinschi2013-DF}, thus the CFO film (bulk lattice constant $a$ = 8.39 $\AA$) was subject to a significant compressive strain of -6.8\% ($a_{\rm SrTiO_3} = 3.905$\AA). In the second study the CFO films were grown on a MgO(100) substrate ($a_{MgO}$ =4.21 \AA)  at 650~$^{\circ}$C leading to a lattice mismatch of only 
0.36\%~\cite{Zviagin2016-DF-both}.

\subsubsection{\label{subsec:optgw}Optical spectrum: IP and $G_{0}W_{0}$} 

The IP spectra with the different starting functionals are plotted in 
Fig.~\ref{fig:opt-abs-spectra}. The analysis is performed with $U_{\rm{Co}}$=$U_{\rm{Fe}}$= 4 eV for PBE+$U$ 
and $U_{\rm{Co}}$=$U_{\rm{Fe}}$= 3 eV for SCAN+$U$ (the results for other $U$ values are given in the supplementary information, Fig.~S6).

The onset of the imaginary part of the optical spectrum is 
at 1.38 eV (PBE+$U$), 1.69~eV (SCAN+$U$), and 2.02~eV (HSE06) and reflects the Kohn-Sham band gap [see Table~\ref{tab:table1}].
For PBE+$U$ ($U_{\rm{Co}}$=$U_{\rm{Fe}}$=4 eV), 
a shoulder is observed at 2.79 eV followed by two peaks at 3.67 and 4.74 eV.
With SCAN+$U$ ($U_{\rm{Co}}$=$U_{\rm{Fe}}$= 3 eV), the shoulder is located at 3.21 eV, and the 
two peaks are at 4.25 and 5.71 eV. Both semi-local functionals 
reproduce the shape of the experimental spectra \wrt{} the spectral features,
 and the intensity of the peaks, but the peak positions are at slightly higher energies than in the experiment. 
With HSE06, the two peaks are further shifted 
to 5.03 and 6.64 eV, compared to the experimental values at 3.50 and 5.0~eV.

Upon including quasiparticle corrections within the $G_{0}W_{0}$ approximation, 
the Im[$\epsilon$($\omega$)] spectrum
is blue shifted to higher energies \wrt{} the IP spectrum for all functionals, but the spectral features from the IP picture are retained to a large extent. The magnitude of the blue shift at the onset for the independent quasiparticle approximation (IQPA) spectrum decreases from 0.48~eV (PBE+$U$) to 0.34~eV (SCAN+$U$) and 0.20 eV (HSE06), reflecting an improved screening effect at the starting DFT level. The prominent shoulder at around 3.5~eV emerges only after including quasiparticle corrections, but, interestingly, it is nearly quenched for PBE+$U$ and SCAN+$U$.

The macroscopic static electronic dielectric constant, $\epsilon_{\infty}$ = Re [$\epsilon$($\omega$ = 0)] in the IP picture is 6.42, 5.68, and 5.08 with PBE+$U$ ($U_{\rm{Co}}$=$U_{\rm{Fe}}$=~4 eV), SCAN+$U$ ($U_{\rm{Co}}$=$U_{\rm{Fe}}$=~3 eV) and HSE06, respectively, where SCAN+$U$ ($U_{\rm{Co}}$=$U_{\rm{Fe}}$=~3 eV) renders good agreement with the experiment at 6~eV~\cite{Zviagin2016-DF-both}. Upon including QP effects, $\epsilon_{\infty}$ decreases to 5.68, 4.99, and 4.62 with PBE+$U$, SCAN+$U$, and HSE06, respectively.
The first peak of Re[$\epsilon$($\omega$)] within IP ($G_{0}W_{0}$) is at 2.36 (2.82), 2.70 (3.20) and 3.70 (4.13)~eV with PBE+$U$, SCAN+$U$, and HSE06, respectively, these values are higher compared to the experimental value of 1.38 eV~\cite{Zviagin2016-DF-both} and 1.95~eV~\cite{Himcinschi2013-DF}.

\subsubsection{\label{subsec:optbse}Optical spectrum including excitonic effects}

Taking into account electron-hole interactions by solving  BSE leads to a significant spectral weight redistribution of the Im[$\epsilon$($\omega$)] spectrum [black solid line in Fig.~\ref{fig:opt-abs-spectra} (d, e, f), see also Fig. S6 in supplementary information for other $U$ values \wrt{} the IP and IQPA spectra].    
The onset of the spectrum is at around 1.45~eV (PBE+$U$), 1.50~eV (SCAN+$U$), and 1.61~eV (HSE06), which is in good agreement with the experimental onset at 1.50~eV. The shoulder at 2.20~eV (PBE+$U$), 2.41~eV (SCAN+$U$), and 2.46 eV (HSE06) corresponds to the broad shoulder at around 2.5~eV in the experimental spectrum. This is followed by a two-peak feature at 2.85 and 3.70~eV (PBE+$U$), 3.1 and 4.0~eV (SCAN+$U$), and 3.49 and 4.62~eV (HSE06), which becomes prominent only after including the excitonic effects and corresponds to the first broad peak at 3.5~eV  (brown solid line, Exp.1~\cite{Zviagin2016-DF-both}).  The energetic position of the second peak in the Im[$\epsilon$($\omega$)] is at around 4.8~eV (PBE+$U$), 5.2~eV (SCAN+$U$), and 5.5~eV (HSE06) compared to the experimental value of 5.0~eV. 
Overall SCAN+$U$ exhibits the best agreement with the spectrum of Zviagin \etal{}~\cite{Zviagin2016-DF-both} with respect to the onset and the position and intensity of the shoulder, as well as the overall shape of the spectrum, in particular at higher energies, underlining the importance of the excitonic effects.

Similarly, the Re[$\epsilon$($\omega$)] spectra are redshifted to lower energies with respect to the IP and $G_{0}W_{0}$ spectra for all functionals upon inclusion of excitonic effects. 
The macroscopic static electronic dielectric constant is 5.08 (PBE+$U$), 5.11 (SCAN+$U$), and, 4.66 (HSE06), which is lower than the experimental value of 6.0~\cite{Himcinschi2013-DF}. 
Furthermore, the first peak in the theoretical Re[$\epsilon$($\omega$)] spectrum is observed at 1.99~eV (PBE+$U$), 2.28~eV (SCAN+$U$) and 2.00~eV (HSE06) compared to the experimental peak at 1.95 eV~\cite{Himcinschi2013-DF}.
Only HSE06 renders a shoulder at 2.92  for the $G_{0}W_{0}+$BSE spectrum, corresponding to the experimental shoulder at 2.90 eV~\cite{Himcinschi2013-DF}. 

We have also calculated the binding energy ($E_\textrm{b}$) of the first exciton in the $G_{0}W_{0}+$BSE spectrum~\cite{Baldini2017-BE-ex}: 0.61 eV (PBE+$U$), 0.59 eV (SCAN+$U$), and 0.85 eV (HSE06).  While to our knowledge there is no experimental report for the exciton binding energy of CFO, the obtained values are comparable to the ones for other related oxides such as SrTiO$_3$ (0.25~eV with SCAN)~\cite{Begum2019-p1} and MgO (0.59~eV with HSE06)~\cite{Begum2021-p2} and SrZrO$_{3}$ (0.31~eV with PBE+$U$)~\cite{Varrassi2021-mBSE}.

To summarize, the overall shape of the optical spectra after BSE does not exhibit a strong dependence on the starting ground state exchange-correlation functional, consistent with previous findings that inclusion of quasiparticle and excitonic effects reduces the dependence on starting exchange-correlation functional~\cite{Begum2019-p1,Begum2021-p2}. Among the three functionals, our BSE calculation starting with SCAN+$U$ renders the best agreement with experiment \wrt{} the onset of the spectrum, and energetic position and intensity of the shoulder and peaks. Overall, our calculations indicate the inclusion of excitonic effects is essential for the investigation of the optical and absorption spectrum of \cofetof.

\subsubsection{\label{subsubsec:abs-spectra}Absorption coefficient spectrum} 
In addition to the optical spectrum in Fig.~\ref{fig:opt-abs-spectra}~(g-l), we have also plotted the absorption coefficient $\alpha(\omega)$, calculated as~\cite{snir2020operando,piccinin2019band,wu2021electronic}:
\begin{equation}
\alpha(\omega) = \dfrac{4\pi}{\lambda}\kappa.
\label{eq-3}
\end{equation}
Here $\lambda$ is the wavelength of incident radiation ($\lambda = 1.24 \times 10^{-6}$ eV.m) and
$\kappa$ is the imaginary part of the refractive index which is related to the real 
Re[$\epsilon$($\omega$)] and imaginary Im[$\epsilon$($\omega$)] part 
of the dielectric function:
\begin{equation}
Re[\epsilon(\omega)] + Im[\epsilon(\omega)] = (n+i\kappa)^2,
\label{eq-4}
\end{equation}
where $n$ is the real part of the refractive index. 
By combining Equations~\ref{eq-3} and \ref{eq-4}, the absorption coefficient is defined as:
\begin{equation}
\alpha(\omega) = \dfrac{4\pi}{\lambda\sqrt{2}}\sqrt{-Re[\epsilon(\omega)]+\sqrt{Re^{2}[\epsilon(\omega)]+Im^{2}[\epsilon(\omega)]}}.
\label{eq-5}
\end{equation}

From Fig.~\ref{fig:opt-abs-spectra}, the onset of $G_{0}W_{0}+$BSE spectrum for the three exchange correlation functionals is 1.45 (PBE+$U$), 1.50 (SCAN+$U$) and 1.61~eV (HSE06), in good correspondence with the measured onset at 1.55~eV (obtained for epitaxial CFO films  grown at 690$^{\circ}$ on MgAl$_{2}$O$_{4}$ ($a$ = 8.08~\AA\ ) substrate with  3.5\% compressive strain)~\cite{Holinsworth2013-opt-gap}. 
The shoulders in the spectra are located at 2.9 and 4.1 eV (PBE+$U$), 2.47 and 4.3 eV (SCAN+$U$), and 2.6 and 4.7 eV (HSE06), in agreement with the two broad experimental shoulders at around 2.6 eV and 4.5 eV.
In general, both HSE06 and SCAN+$U$ render excellent agreement with the measured absorption coefficient spectrum after taking into account the excitonic effects.

We further analyze the oscillator strengths obtained from the BSE calculations [see Fig.~\ref{fig:opt-abs-spectra}~(g-l)] which indicate the probability of excitation at the corresponding energy. In general, the first excitation with a non-zero oscillator strength is interpreted as the optical band gap (the lowest threshold for optical transitions). 
From Fig.~\ref{fig:opt-abs-spectra} we observe the first optically allowed transition marked as 1  at the onset of the spectra at 1.45~eV (PBE+$U$), 1.50~eV (SCAN+$U$) and 1.61 eV (HSE06).
Oscillator strengths with high intensity are found at around 2.0~eV (marked as 2 and 3 in Fig.~\ref{fig:opt-abs-spectra}) for all the functionals, at 3.5~eV (marked as 4 for PBE+$U$, and 4 and 5 for SCAN+$U$ and HSE06) and at around 5.0~eV (marked as 6 for SCAN+$U$ and HSE06). These are in good agreement with the reported optical transitions from ellipsometry measurements for a single crystal of Co$_{1.04\pm 0.05}$Fe$_{1.96 \pm 0.05}$O$_{4}$~\cite{martens1982-DF-CFO} at 2.0, 3.4 and 4.9~eV. From magneto-optical Kerr spectroscopy transitions were reported at 1.82, 2.21, 2.60, 3.55, and 4.0 eV~\cite{Fontijn1999-off-diagonal-DF}, and at 1.78, 2.05, 2.67, 3.6, 4.3 and 4.7~eV~\cite{Himcinschi2013-DF}.
The projected band structure in Fig. S4 indicates that the first transition is from the highest occupied band in the minority spin channel comprised of Co 3$d$ states hybridized with O 2$p$ states to the bottom of the CB which is dominated by 3$d$ states of octahedral Fe.
This suggests that the first allowed transition has a mixed Mott-Hubbard and charge transfer character. 
The transition at around 2~eV stems from Co $3d$ $\longrightarrow$ at the top of the valence band to Fe$^{\rm{Oct}}_{t_{2g}}$ at the bottom of the conduction band, as shown in the projected DOS (Fig.~\ref{fig:PDOS}) and band structure Fig. S4 in SI.

As mentioned already in the introduction, the reported experimental direct optical gaps show a wide range between $\sim$ 0.55 - 4.1~eV. This broad variation can be attributed to the effect of temperature, size, crystallinity, degree of inversion, and shape of the samples.
For the fully inverse bulk spinel our $G_{0}W_{0}$+BSE calculations indicate an optical gap of 1.50~eV (SCAN+$U$) and 1.61~eV (HSE06) in agreement with measured values of 1.65~\cite{Singh2018-opt-gap} and 1.58~eV~\cite{Sharma2014-opt-gap}.
Further optical transitions are at around 2.0, 3.5, and 5.0~eV in agreement with the measurements~\cite{martens1982-DF-CFO}. 

\begin{figure*}[!htbp]
\includegraphics[width=1.0\textwidth]{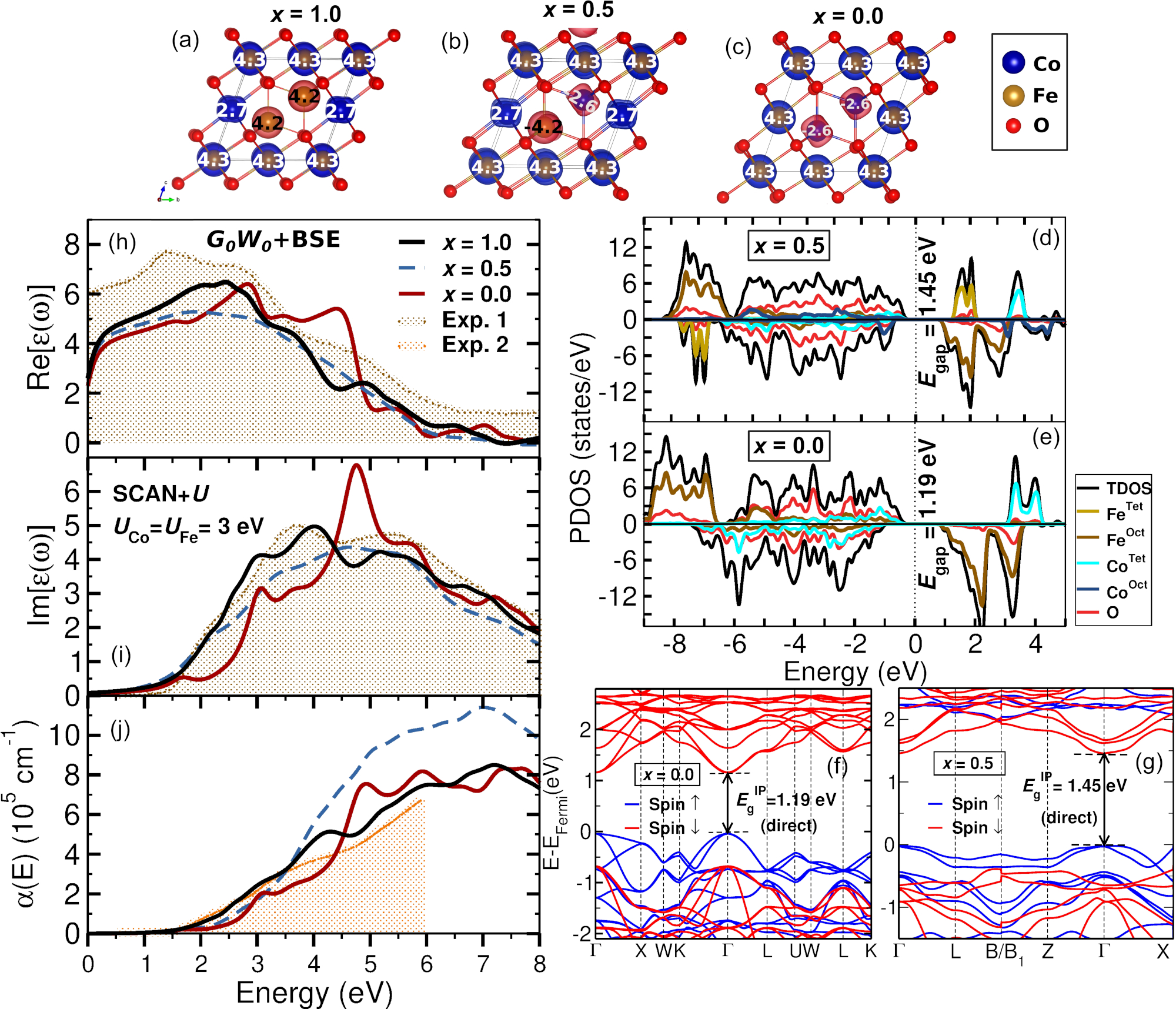} 
\caption{\label{fig:INV} Impact of cation distribution on the magnetic, electronic, and optical properties of \xcfo{} obtained with SCAN+$U$ ($U_{\rm{Co}}$=$U_{\rm{Fe}}$=~3~eV):
 spin density of (a) inverse ($x$ = 1.0) (b) half ($x$ = 0.5) and (c) normal spinel ($x$ = 0.0). The blue and red colors represent positive and negative spin density, respectively. Additionally, the magnetic moments of cations are given in $\mu_{\rm{B}}$. Projected density of states (PDOS) for (d) $x$ = 0.5 and (d) $x$ = 0.0, the calculated IP band structure for (f) $x$ = 0.5 and (g) $x$ = 0.0 (blue/red lines: majority/minority spin channels).
Optical absorption spectrum: (h) real part Re[$\varepsilon(\omega)$], (i) imaginary part Im[$\varepsilon(\omega)$] of the dielectric function (DF) and (j) absorption coefficient ($\alpha(E)$) for $x$ = 1.0 (black solid line), 0.5 (dashed blue line) and 0.0 (red solid line) within $G_{0}W_{0}$+BSE calculations.
Experimental data are adopted from Exp. 1 \cite{Zviagin2016-DF-both} (brown shadowed area), and Exp. 2 \cite{Holinsworth2013-opt-gap} (orange shadowed area). 
}
\end{figure*}

\subsection{\label{subsec:Inversion}Impact of degree of inversion}

As discussed in the introduction, the distribution of cations at octahedral and tetrahedral sites can impact the structural, electronic and optical properties of spinels. 
In this section, we assess the effect of the degree of inversion on the electronic and optical properties by considering  $x$ = 0.0 (normal spinel) and 0.5 (half inverse spinel) in \xcfo{} using the SCAN+$U$ functional with $U_{\rm{Co}}$=$U_{\rm{Fe}}$= 3~eV.

In the normal spinel ($x$ = 0.0) all the Co cations occupy tetrahedral sites while all Fe ions are located at the octahedral sites. For the partially inverse spinel of $x$ = 0.5, half of the Co cations occupy the tetrahedral sites and the remaining half octahedral sites. For all degrees of inversion modeled here, we used the primitive rhombohedral unit cell including two spinel formula units with 14 atoms.
The fully inverse spinel ($x$ = 1.0) is favored in energy by 0.074 and 0.006~eV/f.u. compared to the half ($x$ = 0.5) and normal ($x$ = 0.0) spinel.
The calculated lattice constants are 8.357 ($x$ = 0.5) and 8.371~\AA{} ($x$~=~0.0) which are 0.21\% and 0.38\% larger than the inverse spinel (8.33~\AA ). The trend is in line with the experimental reports by Venturini~\etal{}~\cite{Venturini2019-INV-CFO} who found 8.384~\AA{} for $x$ = 0.0 and 8.364~\AA{} for $x$ = 1.0), as well as previous DFT calculations with the PBEsol exchange correlation functional and $U=4$~eV for both Fe and Co by Sharma~\etal{}\cite{Sharma2022-INV} who reported that the lattice constant increases from  8.332~\AA ($x$ = 1.0) to 8.358 ($x$ = 0.5) and 8.384~\AA ($x$ = 0.0).
From the calculated magnetic moment, presented in Fig.~\ref{fig:INV} b and c, Co and Fe ions preserve the high spin configuration at octahedral and tetrahedral sites for all degrees of inversion in agreement with previous results~\cite{Hou2010-inversionT}. Due to the different sizes of Co and Fe magnetic moments and the antiparallel orientation at octahedral and tetrahdral sites, the total magnetic moment in the unit cell increases to 5 and 7~\mub\ for $x=0.5$ and $x=0.0$.

The PDOS for $x$~=~0.5, presented in Fig.~\ref{fig:INV} d, shows that the top of the valence band is dominated by O~2$p$ and Co$^{\rm{Tet}}$~3$d$ states in the majority and Co$^{\rm{Oct}}$~3$d$  in the minority spin channel. The bottom of the conduction band is comprised of Fe$^{\rm{Oct}}$~3$d$ and Fe$^{\rm{Tet}}$~3$d$ states in the minority and majority spin channels, respectively.  

For $x$~=~0.0 (Fig.~\ref{fig:INV} e), 
the top of the valence band is dominated by Co$^{\rm{Tet}}$~3$d$ and O~2$p$ states in both spin channels. The bottom of the conduction band is comprised of Fe$^{\rm{Oct}}$~3$d$ (minority) and Co$^{\rm{Tet}}$~3$d$ states (majority spin channel). In both cases, similar to the inverse spinel ($x$ = 1.0) presented in Fig.~\ref{fig:PDOS} d, the VBM is located in the majority and the CBM in the minority spin channel. As shown in Fig.~\ref{fig:INV} f and g, the calculated band gap is direct at the $\Gamma$ point and is reduced to 1.45 ($x$ = 0.5)  and 1.19~eV ($x$ = 0.0), compared to 1.69~eV ($x$ = 1.0).
Upon inclusion of QP corrections (cf. Fig. S8 in SI), the band gaps remain direct at $\Gamma$ point and increase to 1.90 ($x=0.5$) and 1.74 eV ($x=0.0$), in contrast to an indirect band gap of  1.95~eV with VBM along $\Gamma$-Y and CBM at $\Gamma$ for  $x=1.0$ (cf. Fig.~\ref{fig:bstr}~h).
The VBM lies in the majority spin channel for both IP and QP band structures for the cases with reduced degree of inversion in contrast to the completely inverse spinel. 

The calculated real and imaginary of the DF as well as the absorption coefficient after the inclusion of the excitonic effects ($G_{0}W_{0}$+BSE calculation) with SCAN+$U$ ($U_{\rm{Co}}$=$U_{\rm{Fe}}$= 3 eV) for $x$ = 0.0, 0.5 and 1.0 are presented in Fig.~\ref{fig:INV}(g-i). As discussed previously in Section~\ref{subsec:optbse}, the experimental spectrum has a shoulder at 2~eV and two broad peaks with nearly equal intensity at around 3.5 and 5~eV~\cite{Himcinschi2013-DF,Zviagin2016-DF-both}. We note that these studies do not provide information on the degree of inversion of the samples.
In the calculated imaginary part of the optical spectrum for $x=1.0$ the first peak is split into two peaks at 3.1 and 4.0~eV with similar intensity as the experimental spectrum (experiment one is a broad peak around 3.5~eV).
However, as presented in Fig.~\ref{fig:INV}(i), the cation distribution influences the position and intensity of the shoulder and peaks of the optical spectrum. Comparison of the optical spectra for $x$ = 0.0, 0.5, and 1.0 indicates a similar onset around 1.50-1.60~eV.  Upon moving Co$^{2+}$ to the tetrahedral sites ($x$ = 0.5 and 0.0), the intensity of the shoulder at around 2~eV decreases. For $x$ = 0.5, one broad featureless peak is observed at around 3.15 to 5.9~eV. 
In the case of the normal spinel ($x$ = 0.0), the shoulder at 2~eV has the lowest intensity with respect to $x$ = 0.5 and 1.0, followed by two small peaks at 3.0 and 3.6~eV, and an increased intensity of the peak at 4.7~eV.  
By analyzing the calculated optical spectra and oscillator strength (Fig. S7 in SI), an optical band gap (the lowest threshold for optical transitions) of 1.64, 1.57, and 1.50~eV is obtained for $x$ = 0.0, 0.5, and 1.0, respectively. 
Our findings suggest that the optical band gap decreases with decreasing degree of inversion, but overall the differences are small which is consistent with previous experimental and theoretical studies for another spinel, 
ZnFe$_{2}$O$_{4}$~\cite{Granone2018-INV-znfe2o4}. On the other hand, the shape of the spectrum shows substantial differences, which allows us to distinguish between different degrees of inversion in CFO samples.

The spectrum of the real part of DF, presented in Fig.~\ref{fig:INV} h, becomes broader and shifts to higher energies with decreasing degree of inversion. The spectrum shows a peak at 2.8~eV and a shoulder at $3.5$ eV ($x$ = 1.0), a rather featureless broad peak between 1.15 - 3.4~eV ($x$ = 0.5), and two peaks at 2.84 and 4.42~eV ($x$ = 0.0). 
Additionally, the macroscopic static electronic dielectric constant is found to decrease from 5.11 ($x$~=~1.0) to 4.31 ($x=0.5$), and 4.54 ($x=0.0$).

From the analysis of the oscillator strength presented in Fig. S7 in SI, both $x$~=~0.0 and 0.5 show transitions below 1~eV with very small non-zero oscillator strength (see insets in Fig. S7 (g-i)). We note that small polarons, as observed in other transition metal oxides, e.g. Co$_{3}$O$_{4}$ and Fe$_{2}$O$_{3}$~\cite{Smart2019-co3o4-polarons,Lohaus2018-fe2o3-polarons}, tend to have much more pronounced midgap transitions. Fontijn~\etal{}~\cite{Fontijn1999-off-diagonal-DF} proposed that in \cofetof{} the transitions below 1~eV originate from the presence of Co$^{2+}$ cations at tetrahedral sites. This is consistent with the calculated optical spectra and PDOS analysis since these transitions are absent in the fully inverse spinel.

Other optical transitions with high-intensity oscillator strength are marked in Fig. S7 at 3.0, 3.6, and 4.7~eV for $x$ = 0.0 and at 2.0 and 3.2~eV for $x$ = 0.5.   
The absorption coefficient spectra of $x$ = 0.0, 0.5, and 1.0 are presented in Fig.~\ref{fig:INV}~j and compared with the experimental spectrum adopted from \cite{Holinsworth2013-opt-gap}. 
With decreasing degree of inversion the onset of the calculated absorption coefficient spectra increases from 1.50~eV ($x$ = 1.0) to 1.57~eV ($x$ = 0.5) and 1.64~eV ($x$ = 0.0). Moreover, the spectra for different degree of inversion exhibit distinct shapes that may be used as a fingerprint. The best agreement with the experimental spectrum is obtained for the fully inverse structure ($x$ = 1.0).

\section{\label{sec:Summary}Summary}
We have systematically investigated the electronic and optical properties of CoFe$_{2}$O$_{4}$ using different levels of description starting with the independent particle picture, and subsequently including quasiparticle ($G_{0}W_{0}$) correction and excitonic ($G_{0}W_{0}+$BSE) effects. 
Moreover, the effect of different starting exchange-correlation functionals (PBE+$U$, SCAN+$U$, and HSE06) and a variation of Hubbard $U$ term on the electronic and optical properties of CFO was explored. In addition, we investigated the effect of the degree of inversion $x$ on the electronic and optical properties of the \xcfo{} with SCAN+$U$.

The starting exchange-correlation functional has a significant influence on the electronic structure at the IP level, in particular, 
with respect to the size and type of band gap (direct/indirect), and the position of the
VBM and CBM in the minority/majority spin channel.
While an indirect band gap of 0.92 and 2.02~eV is obtained with PBE+$U$ ($U_{\rm{Co}}$=$U_{\rm{Fe}}$= 3~eV) and HSE06 in the minority spin channel with VBM along $\Gamma$-$Y$ and CBM at $\Gamma$, a direct band gap of 1.38 and 1.69~eV is obtained with PBE+$U$ ($U_{\rm{Co}}$=$U_{\rm{Fe}}$=~4 eV), SCAN+$U$ ($U_{\rm{Co}}$=$U_{\rm{Fe}}$= 3 eV) wherein the VBM/CBM is located in the majority/minority spin channel at $\Gamma$. The VBM is predominantly comprised of O~2$p$ and Co~3$d$ states, whereas the CBM consists of Fe$^{\rm{oct}}$~3$d$ states.
However, the deviations between the different starting functionals reduce appreciably after including the quasiparticle effects, similar to previous findings for SrTiO$_{3}$~\cite{Begum2019-p1}. 
Including quasiparticle effects ($G_{0}W_{0}$ ), enhances the band gap to 1.78~eV (PBE+$U$, $U_{\rm{Co}}$= $U_{\rm{Fe}}$=~4 eV), 1.95~eV (SCAN+$U$, $U_{\rm{Co}}$= $U_{\rm{Fe}}$=~3 eV) and 2.17 eV (HSE06) and leads to an indirect band gap for all functionals. Moreover, modification of the band structure beyond a rigid band shift underline the critical contribution of the non-local character of the self-energy term in the $GW$ approximation.

Concerning the optical spectra, the imaginary part of the dielectric function obtained with SCAN+$U$ ($U_{\rm{Co}}$= $U_{\rm{Fe}}$= 3 eV) and HSE06 shows good agreement with the experimental optical spectra \cite{Zviagin2016-DF-both,Himcinschi2013-DF} with respect to the energetic position and intensity of peaks only after including  excitonic effects by solving the Bethe Salpeter equation. Also the absorption coefficient obtained with SCAN+$U$ ($U_{\rm{Co}}$= $U_{\rm{Fe}}$= 3 eV) fits very well the measured one~\cite{Holinsworth2013-opt-gap}.
From the analysis of oscillator strength, the lowest threshold for optical transitions, the optical band gap, is at 1.45, 1.50, and 1.61 eV with PBE+$U$ ($U_{\rm{Co}}$= $U_{\rm{Fe}}$= 4 eV), SCAN+$U$ ($U_{\rm{Co}}$= $U_{\rm{Fe}}$= 3 eV) and HSE06, respectively, close to the experimental optical gap of 1.65~\cite{Singh2018-opt-gap} and 1.58~eV~\cite{Sharma2014-opt-gap}. Moreover, the caclulated spectra
show transitions at $\sim$2, 3.5, and 5 eV, in agreement with the experimental findings~\cite{martens1982-DF-CFO}.

Additionally, we explored the impact of cation distribution at the tetrahedral and octahedral sites on the structural and optical properties of \xcfo{} ($x$ = 0.0, 0.5, and 1.0) with SCAN+$U$ ($U_{\rm{Co}}$=$U_{\rm{Fe}}$= 3 eV). With decreasing  degree of inversion, the lattice constant as well as the total magnetic moment per f.u. increase in agreement with previous theoretical and experimental studies~\cite{Venturini2019-INV-CFO,Hou2010-inversionT}. 
While at the IP/QP level the band gap decreases with the degree of inversion, 1.69/1.96~eV ($x$ = 1.0),  1.45/1.90~eV ($x$ = 0.5) and 1.19/1.74~eV ($x$ = 0.0), the optical gap after including excitonic effects shows a slight increase from 1.50~eV ($x$ = 1.0) to 1.57~eV ($x$ = 0.5) and 1.64~eV ($x$ = 0.0). The presence of Co ions in the tetrahedral sites significantly modifies the overall shape of the spectrum and leads to transitions below 1~eV with very small non-zero oscillator strength which are not present in the fully inverse spinel structure, consistent with previous experimental suggestions~\cite{Holinsworth2013-opt-gap}.

\begin{figure*}[!htp]
\includegraphics[width=0.9\textwidth]{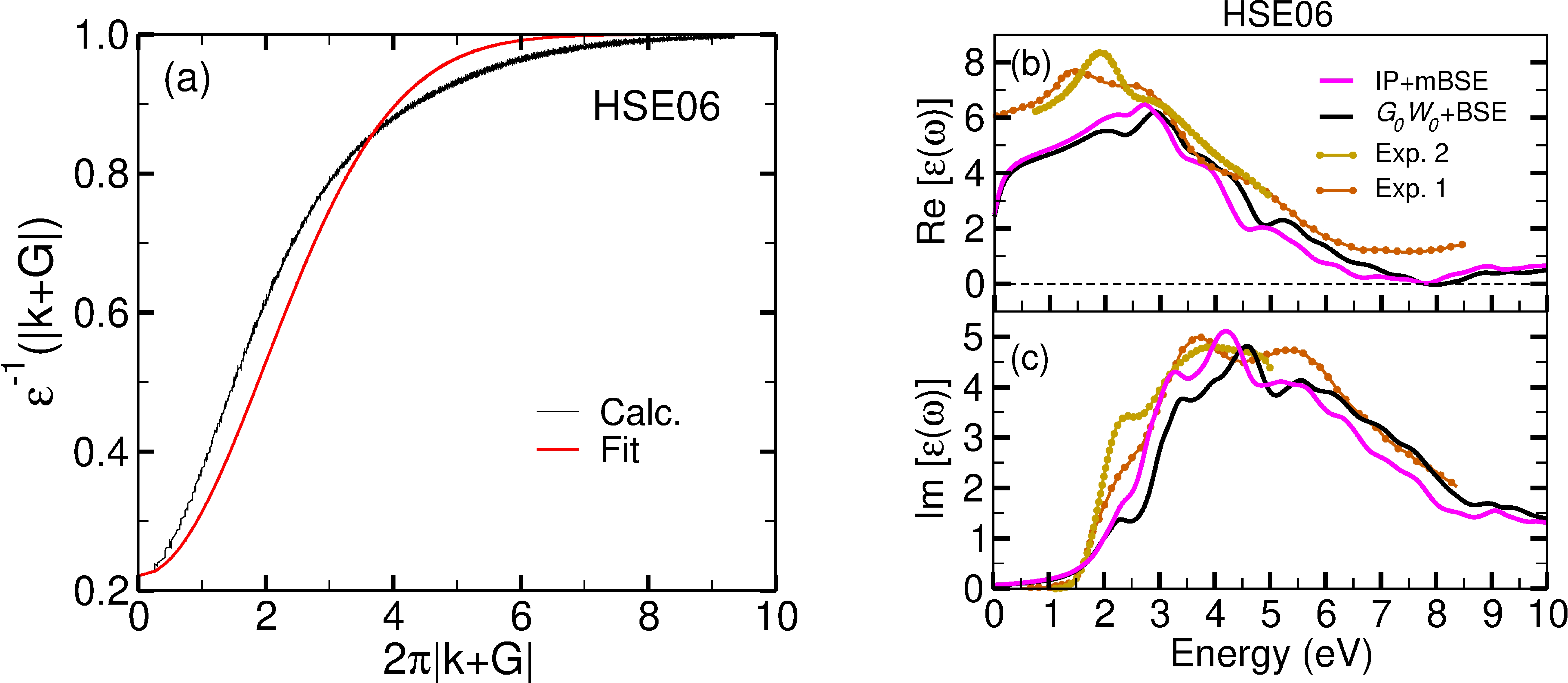}
\caption{(a) Inverse of dielectric function $\varepsilon^{-1}$ from $G_{0}W_{0}$ calculation and 
the corresponding fit according to equation \ref{Eq:mBSE} with HSE06 functional, (b and c)  Comparison of the
real Rm[$\epsilon$($\omega$)] and imaginary Im[$\epsilon$($\omega$)] parts of dielectric function 
using $G_{0}W_{0}+$BSE and model BSE approach. Experimental data are adopted from Exp. 1 \cite{Zviagin2016-DF-both} and Exp. 2 \cite{Himcinschi2013-DF}} 
\label{fig:mBSE}
\end{figure*}

The detailed analysis of the electronic and optical properties of CFO employing DFT calculations and state-of-the-art many-body perturbation theory (MBPT) is useful not only for the interpretation of experimental measurements, but  is also a prerequisite for exploring the incorporation of CFO in heterostructures and nanocomposites in view of carrier separation and reduction of recombination rates.


\appendix
\section{\label{sec:mBSE} Model BSE}

To reduce the computational cost in $G_{0}W_{0}+$BSE calculations, we tested a less computationally demanding approach for the static screening, the model BSE (mBSE) scheme \cite{Bokdam2016-MBSE,Fuchs2008-mBSE,Liu2018-mBSE,Bechstedt1992-mBSE}.

In this approach, the imaginary part of the dielectric constant is replaced by the local model function and is fitted to $G_{0}W_{0}$ calculations using~\cite{Bechstedt1992-mBSE}:

\begin{equation}
\varepsilon^{-1}_{\textbf{k}+\textbf{G}} = 1 - (1-\varepsilon^{-1}_{\infty})e^{\dfrac{-|\textbf{k}+\textbf{G}|^{2}}{4\beta^{2}}} ,
\label{Eq:mBSE}
\end{equation}

where $\beta$ is the range separation parameter, \textbf{G} is the lattice vector, and $\varepsilon_{\infty}$ is the ion-clamped static dielectric function. Here, $\beta$ is obtained by fitting the screened Coulomb kernel diagonal values from the $G_{0}W_{0}$ calculation as shown in Fig.~\ref{fig:mBSE} (a). 
A scissor operator, $\Delta$ is applied to the Kohn-Sham eigenenergies to mimic the QP effect and is defined as a difference between the $G_{0}W_{0}$ and IP band gap. The electron-hole interactions are considered by solving BSE using the Kohn-Sham wave functions.

In the mBSE calculations starting from HSE06, $\beta =$~1.414 and $\varepsilon_{\infty} =$~0.221 (obtained from the fit to the $G_{0}W_{0}$ dielectric function shown in Fig.~\ref{fig:mBSE} (a)) are used as input.
A $\Gamma$-centered 5 $\times$ 5 $\times$ 5 $\mathbf{k}$-mesh is employed for the calculation. 
To converge the electron-hole excitation energy in the range of 0-6~eV, similar to the previous BSE calculations, 24 occupied and 28 unoccupied bands are included in the mBSE calculations. 
As depicted in Fig.~\ref{fig:mBSE} (b and c), the onset of the Im[$\epsilon$($\omega$)] of the model BSE spectrum is at 1.60~eV and is in very good agreement with the $G_{0}W_{0}+$BSE onset at 1.61~eV. 
The first transition at 1.63 eV with a non-zero oscillator strength is in agreement with 1.61~eV from $G_{0}W_{0}+$BSE calculation. 
The feature at around 2.46~eV and the peak 3.49~eV are concurrent with the $G_{0}W_{0}+$BSE spectrum. 
Moreover, the peak at 4.2~eV is in close correspondence with the peak at 4.7~eV. 
An overall good agreement of the model BSE and $G_{0}W_{0}+$BSE spectra is traced back to only small modifications of the HSE06 band structure by including quasiparticle effects beyond the rigid shift, as described in Fig.~\ref{fig:bstr} (f and i). 
We note that starting with PBE+$U$ and SCAN+$U$, we obtain a poor agreement between the model BSE and $G_{0}W_{0}+$BSE spectra [See Fig. S9 in SI]. 
This is attributed to the changes in the electronic structure after $G_{0}W_{0}$, wherein we observed band modifications in the band dispersion.

\begin{acknowledgments}
We acknowledge support by the German Research Foundation
(DFG, Deutsche Forschungsgemeinschaft) within the Collaborative Research
Center TRR247 (Project No. 388390466, Subproject B4), CRC1242 (Project No. 278162697, Subproject No. C02)
and computational time at magnitUDE of the
Center of Computer Science and Simulation (DFG grant
INST 20876/209-1 FUGG,INST 20876/243-1 FUGG).
\end{acknowledgments}



\begin{thebibliography}{81}%
\makeatletter
\providecommand \@ifxundefined [1]{%
 \@ifx{#1\undefined}
}%
\providecommand \@ifnum [1]{%
 \ifnum #1\expandafter \@firstoftwo
 \else \expandafter \@secondoftwo
 \fi
}%
\providecommand \@ifx [1]{%
 \ifx #1\expandafter \@firstoftwo
 \else \expandafter \@secondoftwo
 \fi
}%
\providecommand \natexlab [1]{#1}%
\providecommand \enquote  [1]{``#1''}%
\providecommand \bibnamefont  [1]{#1}%
\providecommand \bibfnamefont [1]{#1}%
\providecommand \citenamefont [1]{#1}%
\providecommand \href@noop [0]{\@secondoftwo}%
\providecommand \href [0]{\begingroup \@sanitize@url \@href}%
\providecommand \@href[1]{\@@startlink{#1}\@@href}%
\providecommand \@@href[1]{\endgroup#1\@@endlink}%
\providecommand \@sanitize@url [0]{\catcode `\\12\catcode `\$12\catcode
  `\&12\catcode `\#12\catcode `\^12\catcode `\_12\catcode `\%12\relax}%
\providecommand \@@startlink[1]{}%
\providecommand \@@endlink[0]{}%
\providecommand \url  [0]{\begingroup\@sanitize@url \@url }%
\providecommand \@url [1]{\endgroup\@href {#1}{\urlprefix }}%
\providecommand \urlprefix  [0]{URL }%
\providecommand \Eprint [0]{\href }%
\providecommand \doibase [0]{https://doi.org/}%
\providecommand \selectlanguage [0]{\@gobble}%
\providecommand \bibinfo  [0]{\@secondoftwo}%
\providecommand \bibfield  [0]{\@secondoftwo}%
\providecommand \translation [1]{[#1]}%
\providecommand \BibitemOpen [0]{}%
\providecommand \bibitemStop [0]{}%
\providecommand \bibitemNoStop [0]{.\EOS\space}%
\providecommand \EOS [0]{\spacefactor3000\relax}%
\providecommand \BibitemShut  [1]{\csname bibitem#1\endcsname}%
\let\auto@bib@innerbib\@empty
\bibitem [{\citenamefont {Dresselhaus}\ and\ \citenamefont
  {Thomas}(2001)}]{dresselhaus2001alternative}%
  \BibitemOpen
  \bibfield  {author} {\bibinfo {author} {\bibfnamefont {M.~S.}\ \bibnamefont
  {Dresselhaus}}\ and\ \bibinfo {author} {\bibfnamefont {I.~L.}\ \bibnamefont
  {Thomas}},\ }\href {https://doi.org/10.1038/35104599} {\bibfield  {journal}
  {\bibinfo  {journal} {Nature}\ }\textbf {\bibinfo {volume} {414}},\ \bibinfo
  {pages} {332} (\bibinfo {year} {2001})}\BibitemShut {NoStop}%
\bibitem [{\citenamefont {Chu}\ and\ \citenamefont
  {Majumdar}(2012)}]{chu2012opportunities}%
  \BibitemOpen
  \bibfield  {author} {\bibinfo {author} {\bibfnamefont {S.}~\bibnamefont
  {Chu}}\ and\ \bibinfo {author} {\bibfnamefont {A.}~\bibnamefont {Majumdar}},\
  }\href {https://doi.org/10.1038/nature11475} {\bibfield  {journal} {\bibinfo
  {journal} {Nature}\ }\textbf {\bibinfo {volume} {488}},\ \bibinfo {pages}
  {294} (\bibinfo {year} {2012})}\BibitemShut {NoStop}%
\bibitem [{\citenamefont {Henrich}\ and\ \citenamefont
  {Cox}(1996)}]{henrich1996surface}%
  \BibitemOpen
  \bibfield  {author} {\bibinfo {author} {\bibfnamefont {V.~E.}\ \bibnamefont
  {Henrich}}\ and\ \bibinfo {author} {\bibfnamefont {P.~A.}\ \bibnamefont
  {Cox}},\ }\href@noop {} {}\ (\bibinfo  {publisher} {Cambridge university
  press},\ \bibinfo {year} {1996})\BibitemShut {NoStop}%
\bibitem [{\citenamefont {Hajiyani}\ and\ \citenamefont
  {Pentcheva}(2018)}]{hajiyani2018surface}%
  \BibitemOpen
  \bibfield  {author} {\bibinfo {author} {\bibfnamefont {H.}~\bibnamefont
  {Hajiyani}}\ and\ \bibinfo {author} {\bibfnamefont {R.}~\bibnamefont
  {Pentcheva}},\ }\href {https://doi.org/10.1021/acscatal.8b00574} {\bibfield
  {journal} {\bibinfo  {journal} {ACS Catal.}\ }\textbf {\bibinfo {volume}
  {8}},\ \bibinfo {pages} {11773} (\bibinfo {year} {2018})}\BibitemShut
  {NoStop}%
\bibitem [{\citenamefont {Peng}\ \emph {et~al.}(2021)\citenamefont {Peng},
  \citenamefont {Hajiyani},\ and\ \citenamefont
  {Pentcheva}}]{peng2021influence}%
  \BibitemOpen
  \bibfield  {author} {\bibinfo {author} {\bibfnamefont {Y.}~\bibnamefont
  {Peng}}, \bibinfo {author} {\bibfnamefont {H.}~\bibnamefont {Hajiyani}},\
  and\ \bibinfo {author} {\bibfnamefont {R.}~\bibnamefont {Pentcheva}},\ }\href
  {https://doi.org/10.1021/acscatal.1c00214} {\bibfield  {journal} {\bibinfo
  {journal} {ACS Catal.}\ }\textbf {\bibinfo {volume} {11}},\ \bibinfo {pages}
  {5601} (\bibinfo {year} {2021})}\BibitemShut {NoStop}%
\bibitem [{\citenamefont {Mulakaluri}\ \emph {et~al.}(2009)\citenamefont
  {Mulakaluri}, \citenamefont {Pentcheva}, \citenamefont {Wieland},
  \citenamefont {Moritz},\ and\ \citenamefont
  {Scheffler}}]{PhysRevLett.103.176102}%
  \BibitemOpen
  \bibfield  {author} {\bibinfo {author} {\bibfnamefont {N.}~\bibnamefont
  {Mulakaluri}}, \bibinfo {author} {\bibfnamefont {R.}~\bibnamefont
  {Pentcheva}}, \bibinfo {author} {\bibfnamefont {M.}~\bibnamefont {Wieland}},
  \bibinfo {author} {\bibfnamefont {W.}~\bibnamefont {Moritz}},\ and\ \bibinfo
  {author} {\bibfnamefont {M.}~\bibnamefont {Scheffler}},\ }\href
  {https://doi.org/10.1103/PhysRevLett.103.176102} {\bibfield  {journal}
  {\bibinfo  {journal} {Phys. Rev. Lett.}\ }\textbf {\bibinfo {volume} {103}},\
  \bibinfo {pages} {176102} (\bibinfo {year} {2009})}\BibitemShut {NoStop}%
\bibitem [{\citenamefont {Chakrapani}\ \emph {et~al.}(2017)\citenamefont
  {Chakrapani}, \citenamefont {Bendt}, \citenamefont {Hajiyani}, \citenamefont
  {Schwarzrock}, \citenamefont {Lunkenbein}, \citenamefont {Salamon},
  \citenamefont {Landers}, \citenamefont {Wende}, \citenamefont {Schl{\"o}gl},
  \citenamefont {Pentcheva}, \citenamefont {Behrens},\ and\ \citenamefont
  {Schulz}}]{chakrapani2017role}%
  \BibitemOpen
  \bibfield  {author} {\bibinfo {author} {\bibfnamefont {K.}~\bibnamefont
  {Chakrapani}}, \bibinfo {author} {\bibfnamefont {G.}~\bibnamefont {Bendt}},
  \bibinfo {author} {\bibfnamefont {H.}~\bibnamefont {Hajiyani}}, \bibinfo
  {author} {\bibfnamefont {I.}~\bibnamefont {Schwarzrock}}, \bibinfo {author}
  {\bibfnamefont {T.}~\bibnamefont {Lunkenbein}}, \bibinfo {author}
  {\bibfnamefont {S.}~\bibnamefont {Salamon}}, \bibinfo {author} {\bibfnamefont
  {J.}~\bibnamefont {Landers}}, \bibinfo {author} {\bibfnamefont
  {H.}~\bibnamefont {Wende}}, \bibinfo {author} {\bibfnamefont
  {R.}~\bibnamefont {Schl{\"o}gl}}, \bibinfo {author} {\bibfnamefont
  {R.}~\bibnamefont {Pentcheva}}, \bibinfo {author} {\bibfnamefont
  {M.}~\bibnamefont {Behrens}},\ and\ \bibinfo {author} {\bibfnamefont
  {S.}~\bibnamefont {Schulz}},\ }\href {https://doi.org/10.1002/cctc.201700376}
  {\bibfield  {journal} {\bibinfo  {journal} {ChemCatChem}\ }\textbf {\bibinfo
  {volume} {9}},\ \bibinfo {pages} {2988} (\bibinfo {year} {2017})}\BibitemShut
  {NoStop}%
\bibitem [{\citenamefont {Schmitz-Antoniak}\ \emph {et~al.}(2013)\citenamefont
  {Schmitz-Antoniak}, \citenamefont {Schmitz}, \citenamefont {Borisov},
  \citenamefont {de~Groot}, \citenamefont {Stienen}, \citenamefont {Warland},
  \citenamefont {Krumme}, \citenamefont {Feyerherm}, \citenamefont {Dudzik},
  \citenamefont {Kleemann},\ and\ \citenamefont {Wende}}]{schmitz2013electric}%
  \BibitemOpen
  \bibfield  {author} {\bibinfo {author} {\bibfnamefont {C.}~\bibnamefont
  {Schmitz-Antoniak}}, \bibinfo {author} {\bibfnamefont {D.}~\bibnamefont
  {Schmitz}}, \bibinfo {author} {\bibfnamefont {P.}~\bibnamefont {Borisov}},
  \bibinfo {author} {\bibfnamefont {F.~M.~F.}\ \bibnamefont {de~Groot}},
  \bibinfo {author} {\bibfnamefont {S.}~\bibnamefont {Stienen}}, \bibinfo
  {author} {\bibfnamefont {A.}~\bibnamefont {Warland}}, \bibinfo {author}
  {\bibfnamefont {B.}~\bibnamefont {Krumme}}, \bibinfo {author} {\bibfnamefont
  {R.}~\bibnamefont {Feyerherm}}, \bibinfo {author} {\bibfnamefont
  {E.}~\bibnamefont {Dudzik}}, \bibinfo {author} {\bibfnamefont
  {W.}~\bibnamefont {Kleemann}},\ and\ \bibinfo {author} {\bibfnamefont
  {H.}~\bibnamefont {Wende}},\ }\href {https://doi.org/10.1038/ncomms3051}
  {\bibfield  {journal} {\bibinfo  {journal} {Nat. Commun.}\ }\textbf {\bibinfo
  {volume} {4}},\ \bibinfo {pages} {2051} (\bibinfo {year} {2013})}\BibitemShut
  {NoStop}%
\bibitem [{\citenamefont {Venturini}\ \emph
  {et~al.}(2019{\natexlab{a}})\citenamefont {Venturini}, \citenamefont
  {Tonelli}, \citenamefont {Wermuth}, \citenamefont {Zampiva}, \citenamefont
  {Arcaro}, \citenamefont {Da~Cas~Viegas},\ and\ \citenamefont
  {Bergmann}}]{Venturini2019-INV}%
  \BibitemOpen
  \bibfield  {author} {\bibinfo {author} {\bibfnamefont {J.}~\bibnamefont
  {Venturini}}, \bibinfo {author} {\bibfnamefont {A.~M.}\ \bibnamefont
  {Tonelli}}, \bibinfo {author} {\bibfnamefont {T.~B.}\ \bibnamefont
  {Wermuth}}, \bibinfo {author} {\bibfnamefont {R.~Y.~S.}\ \bibnamefont
  {Zampiva}}, \bibinfo {author} {\bibfnamefont {S.}~\bibnamefont {Arcaro}},
  \bibinfo {author} {\bibfnamefont {A.}~\bibnamefont {Da~Cas~Viegas}},\ and\
  \bibinfo {author} {\bibfnamefont {C.~P.}\ \bibnamefont {Bergmann}},\ }\href
  {https://www.sciencedirect.com/science/article/pii/S0304885318333432}
  {\bibfield  {journal} {\bibinfo  {journal} {J. Magn. Magn. Mater.}\ }\textbf
  {\bibinfo {volume} {482}},\ \bibinfo {pages} {1} (\bibinfo {year}
  {2019}{\natexlab{a}})}\BibitemShut {NoStop}%
\bibitem [{\citenamefont {Granone}\ \emph {et~al.}(2018)\citenamefont
  {Granone}, \citenamefont {Ulpe}, \citenamefont {Robben}, \citenamefont
  {Klimke}, \citenamefont {Jahns}, \citenamefont {Renz}, \citenamefont
  {Gesing}, \citenamefont {Bredow}, \citenamefont {Dillert},\ and\
  \citenamefont {Bahnemann}}]{Granone2018-INV-znfe2o4}%
  \BibitemOpen
  \bibfield  {author} {\bibinfo {author} {\bibfnamefont {L.~I.}\ \bibnamefont
  {Granone}}, \bibinfo {author} {\bibfnamefont {A.~C.}\ \bibnamefont {Ulpe}},
  \bibinfo {author} {\bibfnamefont {L.}~\bibnamefont {Robben}}, \bibinfo
  {author} {\bibfnamefont {S.}~\bibnamefont {Klimke}}, \bibinfo {author}
  {\bibfnamefont {M.}~\bibnamefont {Jahns}}, \bibinfo {author} {\bibfnamefont
  {F.}~\bibnamefont {Renz}}, \bibinfo {author} {\bibfnamefont {T.~M.}\
  \bibnamefont {Gesing}}, \bibinfo {author} {\bibfnamefont {T.}~\bibnamefont
  {Bredow}}, \bibinfo {author} {\bibfnamefont {R.}~\bibnamefont {Dillert}},\
  and\ \bibinfo {author} {\bibfnamefont {D.~W.}\ \bibnamefont {Bahnemann}},\
  }\href {https://doi.org/10.1039/C8CP05061A} {\bibfield  {journal} {\bibinfo
  {journal} {Phys. Chem. Chem. Phys.}\ }\textbf {\bibinfo {volume} {20}},\
  \bibinfo {pages} {28267} (\bibinfo {year} {2018})}\BibitemShut {NoStop}%
\bibitem [{\citenamefont {Sharma}\ \emph {et~al.}(2022)\citenamefont {Sharma},
  \citenamefont {Calmels}, \citenamefont {Li}, \citenamefont {Barbier},\ and\
  \citenamefont {Arras}}]{Sharma2022-INV}%
  \BibitemOpen
  \bibfield  {author} {\bibinfo {author} {\bibfnamefont {K.}~\bibnamefont
  {Sharma}}, \bibinfo {author} {\bibfnamefont {L.}~\bibnamefont {Calmels}},
  \bibinfo {author} {\bibfnamefont {D.}~\bibnamefont {Li}}, \bibinfo {author}
  {\bibfnamefont {A.}~\bibnamefont {Barbier}},\ and\ \bibinfo {author}
  {\bibfnamefont {R.}~\bibnamefont {Arras}},\ }\href
  {https://doi.org/10.1103/PhysRevMaterials.6.124402} {\bibfield  {journal}
  {\bibinfo  {journal} {Phys. Rev. Mater.}\ }\textbf {\bibinfo {volume} {6}},\
  \bibinfo {pages} {124402} (\bibinfo {year} {2022})}\BibitemShut {NoStop}%
\bibitem [{\citenamefont {Zheng}\ \emph {et~al.}(2004)\citenamefont {Zheng},
  \citenamefont {Wang}, \citenamefont {Lofland}, \citenamefont {Ma},
  \citenamefont {Mohaddes-Ardabili}, \citenamefont {Zhao}, \citenamefont
  {Salamanca-Riba}, \citenamefont {Shinde}, \citenamefont {Ogale},
  \citenamefont {Bai}, \citenamefont {Viehland}, \citenamefont {Jia},
  \citenamefont {Schlom}, \citenamefont {Wuttig}, \citenamefont {Roytburd},\
  and\ \citenamefont {Ramesh}}]{zheng2004multiferroic}%
  \BibitemOpen
  \bibfield  {author} {\bibinfo {author} {\bibfnamefont {H.}~\bibnamefont
  {Zheng}}, \bibinfo {author} {\bibfnamefont {J.}~\bibnamefont {Wang}},
  \bibinfo {author} {\bibfnamefont {S.~E.}\ \bibnamefont {Lofland}}, \bibinfo
  {author} {\bibfnamefont {Z.}~\bibnamefont {Ma}}, \bibinfo {author}
  {\bibfnamefont {L.}~\bibnamefont {Mohaddes-Ardabili}}, \bibinfo {author}
  {\bibfnamefont {T.}~\bibnamefont {Zhao}}, \bibinfo {author} {\bibfnamefont
  {L.}~\bibnamefont {Salamanca-Riba}}, \bibinfo {author} {\bibfnamefont
  {S.~R.}\ \bibnamefont {Shinde}}, \bibinfo {author} {\bibfnamefont {S.~B.}\
  \bibnamefont {Ogale}}, \bibinfo {author} {\bibfnamefont {F.}~\bibnamefont
  {Bai}}, \bibinfo {author} {\bibfnamefont {D.}~\bibnamefont {Viehland}},
  \bibinfo {author} {\bibfnamefont {Y.}~\bibnamefont {Jia}}, \bibinfo {author}
  {\bibfnamefont {D.~G.}\ \bibnamefont {Schlom}}, \bibinfo {author}
  {\bibfnamefont {M.}~\bibnamefont {Wuttig}}, \bibinfo {author} {\bibfnamefont
  {A.}~\bibnamefont {Roytburd}},\ and\ \bibinfo {author} {\bibfnamefont
  {R.}~\bibnamefont {Ramesh}},\ }\href
  {https://doi.org/10.1126/science.1094207} {\bibfield  {journal} {\bibinfo
  {journal} {Science}\ }\textbf {\bibinfo {volume} {303}},\ \bibinfo {pages}
  {661} (\bibinfo {year} {2004})}\BibitemShut {NoStop}%
\bibitem [{\citenamefont {Zavaliche}\ \emph {et~al.}(2007)\citenamefont
  {Zavaliche}, \citenamefont {Zhao}, \citenamefont {Zheng}, \citenamefont
  {Straub}, \citenamefont {Cruz}, \citenamefont {Yang}, \citenamefont {Hao},\
  and\ \citenamefont {Ramesh}}]{zavaliche2007electrically}%
  \BibitemOpen
  \bibfield  {author} {\bibinfo {author} {\bibfnamefont {F.}~\bibnamefont
  {Zavaliche}}, \bibinfo {author} {\bibfnamefont {T.}~\bibnamefont {Zhao}},
  \bibinfo {author} {\bibfnamefont {H.}~\bibnamefont {Zheng}}, \bibinfo
  {author} {\bibfnamefont {F.}~\bibnamefont {Straub}}, \bibinfo {author}
  {\bibfnamefont {M.~P.}\ \bibnamefont {Cruz}}, \bibinfo {author}
  {\bibfnamefont {P.-L.}\ \bibnamefont {Yang}}, \bibinfo {author}
  {\bibfnamefont {D.}~\bibnamefont {Hao}},\ and\ \bibinfo {author}
  {\bibfnamefont {R.}~\bibnamefont {Ramesh}},\ }\href
  {https://doi.org/10.1021/nl070465o} {\bibfield  {journal} {\bibinfo
  {journal} {Nano Lett.}\ }\textbf {\bibinfo {volume} {7}},\ \bibinfo {pages}
  {1586} (\bibinfo {year} {2007})}\BibitemShut {NoStop}%
\bibitem [{\citenamefont {Kampermann}\ \emph {et~al.}(2021)\citenamefont
  {Kampermann}, \citenamefont {Klein}, \citenamefont {Korte}, \citenamefont
  {Kowollik}, \citenamefont {Pfingsten}, \citenamefont {Smola}, \citenamefont
  {Saddeler}, \citenamefont {Piotrowiak}, \citenamefont {Salamon},
  \citenamefont {Landers}, \citenamefont {Wende}, \citenamefont {Ludwig},
  \citenamefont {Schulz},\ and\ \citenamefont {Bacher}}]{kampermann2021link}%
  \BibitemOpen
  \bibfield  {author} {\bibinfo {author} {\bibfnamefont {L.}~\bibnamefont
  {Kampermann}}, \bibinfo {author} {\bibfnamefont {J.}~\bibnamefont {Klein}},
  \bibinfo {author} {\bibfnamefont {J.}~\bibnamefont {Korte}}, \bibinfo
  {author} {\bibfnamefont {O.}~\bibnamefont {Kowollik}}, \bibinfo {author}
  {\bibfnamefont {O.}~\bibnamefont {Pfingsten}}, \bibinfo {author}
  {\bibfnamefont {T.}~\bibnamefont {Smola}}, \bibinfo {author} {\bibfnamefont
  {S.}~\bibnamefont {Saddeler}}, \bibinfo {author} {\bibfnamefont {T.~H.}\
  \bibnamefont {Piotrowiak}}, \bibinfo {author} {\bibfnamefont
  {S.}~\bibnamefont {Salamon}}, \bibinfo {author} {\bibfnamefont
  {J.}~\bibnamefont {Landers}}, \bibinfo {author} {\bibfnamefont
  {H.}~\bibnamefont {Wende}}, \bibinfo {author} {\bibfnamefont
  {A.}~\bibnamefont {Ludwig}}, \bibinfo {author} {\bibfnamefont
  {S.}~\bibnamefont {Schulz}},\ and\ \bibinfo {author} {\bibfnamefont
  {G.}~\bibnamefont {Bacher}},\ }\href
  {https://doi.org/10.1021/acs.jpcc.0c11277} {\bibfield  {journal} {\bibinfo
  {journal} {J. Phys. Chem.}\ }\textbf {\bibinfo {volume} {125}},\ \bibinfo
  {pages} {14356} (\bibinfo {year} {2021})}\BibitemShut {NoStop}%
\bibitem [{\citenamefont {Holinsworth}\ \emph {et~al.}(2013)\citenamefont
  {Holinsworth}, \citenamefont {Mazumdar}, \citenamefont {Sims}, \citenamefont
  {Sun}, \citenamefont {Yurtisigi}, \citenamefont {Sarker}, \citenamefont
  {Gupta}, \citenamefont {Butler},\ and\ \citenamefont
  {Musfeldt}}]{Holinsworth2013-opt-gap}%
  \BibitemOpen
  \bibfield  {author} {\bibinfo {author} {\bibfnamefont {B.~S.}\ \bibnamefont
  {Holinsworth}}, \bibinfo {author} {\bibfnamefont {D.}~\bibnamefont
  {Mazumdar}}, \bibinfo {author} {\bibfnamefont {H.}~\bibnamefont {Sims}},
  \bibinfo {author} {\bibfnamefont {Q.-C.}\ \bibnamefont {Sun}}, \bibinfo
  {author} {\bibfnamefont {M.~K.}\ \bibnamefont {Yurtisigi}}, \bibinfo {author}
  {\bibfnamefont {S.~K.}\ \bibnamefont {Sarker}}, \bibinfo {author}
  {\bibfnamefont {A.}~\bibnamefont {Gupta}}, \bibinfo {author} {\bibfnamefont
  {W.~H.}\ \bibnamefont {Butler}},\ and\ \bibinfo {author} {\bibfnamefont
  {J.~L.}\ \bibnamefont {Musfeldt}},\ }\href
  {https://doi.org/10.1063/1.4818315} {\bibfield  {journal} {\bibinfo
  {journal} {Appl. Phys. Lett.}\ }\textbf {\bibinfo {volume} {103}},\ \bibinfo
  {pages} {082406} (\bibinfo {year} {2013})}\BibitemShut {NoStop}%
\bibitem [{\citenamefont {Himcinschi}\ \emph {et~al.}(2013)\citenamefont
  {Himcinschi}, \citenamefont {Vrejoiu}, \citenamefont {Salvan}, \citenamefont
  {Fronk}, \citenamefont {Talkenberger}, \citenamefont {Zahn}, \citenamefont
  {Rafaja},\ and\ \citenamefont {Kortus}}]{Himcinschi2013-DF}%
  \BibitemOpen
  \bibfield  {author} {\bibinfo {author} {\bibfnamefont {C.}~\bibnamefont
  {Himcinschi}}, \bibinfo {author} {\bibfnamefont {I.}~\bibnamefont {Vrejoiu}},
  \bibinfo {author} {\bibfnamefont {G.}~\bibnamefont {Salvan}}, \bibinfo
  {author} {\bibfnamefont {M.}~\bibnamefont {Fronk}}, \bibinfo {author}
  {\bibfnamefont {A.}~\bibnamefont {Talkenberger}}, \bibinfo {author}
  {\bibfnamefont {D.~R.~T.}\ \bibnamefont {Zahn}}, \bibinfo {author}
  {\bibfnamefont {D.}~\bibnamefont {Rafaja}},\ and\ \bibinfo {author}
  {\bibfnamefont {J.}~\bibnamefont {Kortus}},\ }\href
  {https://doi.org/10.1063/1.4792749} {\bibfield  {journal} {\bibinfo
  {journal} {J. Appl. Phys.}\ }\textbf {\bibinfo {volume} {113}},\ \bibinfo
  {pages} {084101} (\bibinfo {year} {2013})}\BibitemShut {NoStop}%
\bibitem [{\citenamefont {Kalam}\ \emph {et~al.}(2018)\citenamefont {Kalam},
  \citenamefont {Al-Sehemi}, \citenamefont {Assiri}, \citenamefont {Du},
  \citenamefont {Ahmad}, \citenamefont {Ahmad},\ and\ \citenamefont
  {Pannipara}}]{Kalam2018-opt-gap}%
  \BibitemOpen
  \bibfield  {author} {\bibinfo {author} {\bibfnamefont {A.}~\bibnamefont
  {Kalam}}, \bibinfo {author} {\bibfnamefont {A.~G.}\ \bibnamefont
  {Al-Sehemi}}, \bibinfo {author} {\bibfnamefont {M.}~\bibnamefont {Assiri}},
  \bibinfo {author} {\bibfnamefont {G.}~\bibnamefont {Du}}, \bibinfo {author}
  {\bibfnamefont {T.}~\bibnamefont {Ahmad}}, \bibinfo {author} {\bibfnamefont
  {I.}~\bibnamefont {Ahmad}},\ and\ \bibinfo {author} {\bibfnamefont
  {M.}~\bibnamefont {Pannipara}},\ }\href
  {https://doi.org/10.1016/j.rinp.2018.01.045} {\bibfield  {journal} {\bibinfo
  {journal} {Results Phys.}\ }\textbf {\bibinfo {volume} {8}},\ \bibinfo
  {pages} {1046} (\bibinfo {year} {2018})}\BibitemShut {NoStop}%
\bibitem [{\citenamefont {Ravindra}\ \emph {et~al.}(2012)\citenamefont
  {Ravindra}, \citenamefont {Padhan},\ and\ \citenamefont
  {Prellier}}]{Ravindra2012-opt-gap}%
  \BibitemOpen
  \bibfield  {author} {\bibinfo {author} {\bibfnamefont {A.~V.}\ \bibnamefont
  {Ravindra}}, \bibinfo {author} {\bibfnamefont {P.}~\bibnamefont {Padhan}},\
  and\ \bibinfo {author} {\bibfnamefont {W.}~\bibnamefont {Prellier}},\ }\href
  {https://doi.org/10.1063/1.4759001} {\bibfield  {journal} {\bibinfo
  {journal} {Appl. Phys. Lett.}\ }\textbf {\bibinfo {volume} {101}},\ \bibinfo
  {pages} {161902} (\bibinfo {year} {2012})}\BibitemShut {NoStop}%
\bibitem [{\citenamefont {Dileep}\ \emph {et~al.}(2014)\citenamefont {Dileep},
  \citenamefont {Loukya}, \citenamefont {Pachauri}, \citenamefont {Gupta},\
  and\ \citenamefont {Datta}}]{Dileep2014-bstr-mBJLDA}%
  \BibitemOpen
  \bibfield  {author} {\bibinfo {author} {\bibfnamefont {K.}~\bibnamefont
  {Dileep}}, \bibinfo {author} {\bibfnamefont {B.}~\bibnamefont {Loukya}},
  \bibinfo {author} {\bibfnamefont {N.}~\bibnamefont {Pachauri}}, \bibinfo
  {author} {\bibfnamefont {A.}~\bibnamefont {Gupta}},\ and\ \bibinfo {author}
  {\bibfnamefont {R.}~\bibnamefont {Datta}},\ }\href
  {https://doi.org/10.1063/1.4895059} {\bibfield  {journal} {\bibinfo
  {journal} {J. Appl. Phys.}\ }\textbf {\bibinfo {volume} {116}},\ \bibinfo
  {pages} {103505} (\bibinfo {year} {2014})}\BibitemShut {NoStop}%
\bibitem [{\citenamefont {Singh}\ and\ \citenamefont
  {Khare}(2018)}]{Singh2018-opt-gap}%
  \BibitemOpen
  \bibfield  {author} {\bibinfo {author} {\bibfnamefont {S.}~\bibnamefont
  {Singh}}\ and\ \bibinfo {author} {\bibfnamefont {N.}~\bibnamefont {Khare}},\
  }\href {https://doi.org/10.1038/s41598-018-24947-2} {\bibfield  {journal}
  {\bibinfo  {journal} {Sci. Rep.}\ }\textbf {\bibinfo {volume} {8}},\ \bibinfo
  {pages} {6522} (\bibinfo {year} {2018})}\BibitemShut {NoStop}%
\bibitem [{\citenamefont {Sharma}\ and\ \citenamefont
  {Khare}(2014)}]{Sharma2014-opt-gap}%
  \BibitemOpen
  \bibfield  {author} {\bibinfo {author} {\bibfnamefont {D.}~\bibnamefont
  {Sharma}}\ and\ \bibinfo {author} {\bibfnamefont {N.}~\bibnamefont {Khare}},\
  }\href {https://doi.org/10.1063/1.4890863} {\bibfield  {journal} {\bibinfo
  {journal} {Appl. Phys. Lett.}\ }\textbf {\bibinfo {volume} {105}},\ \bibinfo
  {pages} {032404} (\bibinfo {year} {2014})}\BibitemShut {NoStop}%
\bibitem [{\citenamefont {Singh}\ \emph
  {et~al.}(2020{\natexlab{a}})\citenamefont {Singh}, \citenamefont {Park},
  \citenamefont {Singh}, \citenamefont {Kim}, \citenamefont {Lim},
  \citenamefont {Kumar}, \citenamefont {Kim}, \citenamefont {Lee},\ and\
  \citenamefont {Chae}}]{Singh2020-opt-gap}%
  \BibitemOpen
  \bibfield  {author} {\bibinfo {author} {\bibfnamefont {J.~P.}\ \bibnamefont
  {Singh}}, \bibinfo {author} {\bibfnamefont {J.~Y.}\ \bibnamefont {Park}},
  \bibinfo {author} {\bibfnamefont {V.}~\bibnamefont {Singh}}, \bibinfo
  {author} {\bibfnamefont {S.~H.}\ \bibnamefont {Kim}}, \bibinfo {author}
  {\bibfnamefont {W.~C.}\ \bibnamefont {Lim}}, \bibinfo {author} {\bibfnamefont
  {H.}~\bibnamefont {Kumar}}, \bibinfo {author} {\bibfnamefont {Y.~H.}\
  \bibnamefont {Kim}}, \bibinfo {author} {\bibfnamefont {S.}~\bibnamefont
  {Lee}},\ and\ \bibinfo {author} {\bibfnamefont {K.~H.}\ \bibnamefont
  {Chae}},\ }\href {https://doi.org/10.1039/D0RA01653E} {\bibfield  {journal}
  {\bibinfo  {journal} {RSC Adv.}\ }\textbf {\bibinfo {volume} {10}},\ \bibinfo
  {pages} {21259} (\bibinfo {year} {2020}{\natexlab{a}})}\BibitemShut {NoStop}%
\bibitem [{\citenamefont {Fontijn}\ \emph {et~al.}(1999)\citenamefont
  {Fontijn}, \citenamefont {van~der Zaag}, \citenamefont {Feiner},
  \citenamefont {Metselaar},\ and\ \citenamefont
  {Devillers}}]{Fontijn1999-off-diagonal-DF}%
  \BibitemOpen
  \bibfield  {author} {\bibinfo {author} {\bibfnamefont {W.~F.~J.}\
  \bibnamefont {Fontijn}}, \bibinfo {author} {\bibfnamefont {P.~J.}\
  \bibnamefont {van~der Zaag}}, \bibinfo {author} {\bibfnamefont {L.~F.}\
  \bibnamefont {Feiner}}, \bibinfo {author} {\bibfnamefont {R.}~\bibnamefont
  {Metselaar}},\ and\ \bibinfo {author} {\bibfnamefont {M.~A.~C.}\ \bibnamefont
  {Devillers}},\ }\href {https://doi.org/10.1063/1.369091} {\bibfield
  {journal} {\bibinfo  {journal} {J. Appl. Phys.}\ }\textbf {\bibinfo {volume}
  {85}},\ \bibinfo {pages} {5100} (\bibinfo {year} {1999})}\BibitemShut
  {NoStop}%
\bibitem [{\citenamefont {Curtarolo}\ \emph {et~al.}(2012)\citenamefont
  {Curtarolo}, \citenamefont {Setyawan}, \citenamefont {Hart}, \citenamefont
  {Jahnatek}, \citenamefont {Chepulskii}, \citenamefont {Taylor}, \citenamefont
  {Wang}, \citenamefont {Xue}, \citenamefont {Yang}, \citenamefont {Levy},
  \citenamefont {Mehl}, \citenamefont {Stokes}, \citenamefont {Demchenko},\
  and\ \citenamefont {Morgan}}]{Curtarolo2012-AFLOW}%
  \BibitemOpen
  \bibfield  {author} {\bibinfo {author} {\bibfnamefont {S.}~\bibnamefont
  {Curtarolo}}, \bibinfo {author} {\bibfnamefont {W.}~\bibnamefont {Setyawan}},
  \bibinfo {author} {\bibfnamefont {G.~L.}\ \bibnamefont {Hart}}, \bibinfo
  {author} {\bibfnamefont {M.}~\bibnamefont {Jahnatek}}, \bibinfo {author}
  {\bibfnamefont {R.~V.}\ \bibnamefont {Chepulskii}}, \bibinfo {author}
  {\bibfnamefont {R.~H.}\ \bibnamefont {Taylor}}, \bibinfo {author}
  {\bibfnamefont {S.}~\bibnamefont {Wang}}, \bibinfo {author} {\bibfnamefont
  {J.}~\bibnamefont {Xue}}, \bibinfo {author} {\bibfnamefont {K.}~\bibnamefont
  {Yang}}, \bibinfo {author} {\bibfnamefont {O.}~\bibnamefont {Levy}}, \bibinfo
  {author} {\bibfnamefont {M.~J.}\ \bibnamefont {Mehl}}, \bibinfo {author}
  {\bibfnamefont {H.~T.}\ \bibnamefont {Stokes}}, \bibinfo {author}
  {\bibfnamefont {D.~O.}\ \bibnamefont {Demchenko}},\ and\ \bibinfo {author}
  {\bibfnamefont {D.}~\bibnamefont {Morgan}},\ }\href
  {https://www.sciencedirect.com/science/article/pii/S0927025612000717}
  {\bibfield  {journal} {\bibinfo  {journal} {Comput. Mater. Sci.}\ }\textbf
  {\bibinfo {volume} {58}},\ \bibinfo {pages} {218} (\bibinfo {year}
  {2012})}\BibitemShut {NoStop}%
\bibitem [{\citenamefont {Fritsch}\ and\ \citenamefont
  {Ederer}(2010)}]{Fritsch2010-DFT}%
  \BibitemOpen
  \bibfield  {author} {\bibinfo {author} {\bibfnamefont {D.}~\bibnamefont
  {Fritsch}}\ and\ \bibinfo {author} {\bibfnamefont {C.}~\bibnamefont
  {Ederer}},\ }\href {https://doi.org/10.1103/PhysRevB.82.104117} {\bibfield
  {journal} {\bibinfo  {journal} {Phys. Rev. B}\ }\textbf {\bibinfo {volume}
  {82}},\ \bibinfo {pages} {104117} (\bibinfo {year} {2010})}\BibitemShut
  {NoStop}%
\bibitem [{\citenamefont {Lukashev}\ \emph {et~al.}(2013)\citenamefont
  {Lukashev}, \citenamefont {Burton}, \citenamefont {Smogunov}, \citenamefont
  {Velev},\ and\ \citenamefont {Tsymbal}}]{Lukashev2013-DFT-gap}%
  \BibitemOpen
  \bibfield  {author} {\bibinfo {author} {\bibfnamefont {P.~V.}\ \bibnamefont
  {Lukashev}}, \bibinfo {author} {\bibfnamefont {J.~D.}\ \bibnamefont
  {Burton}}, \bibinfo {author} {\bibfnamefont {A.}~\bibnamefont {Smogunov}},
  \bibinfo {author} {\bibfnamefont {J.~P.}\ \bibnamefont {Velev}},\ and\
  \bibinfo {author} {\bibfnamefont {E.~Y.}\ \bibnamefont {Tsymbal}},\ }\href
  {https://doi.org/10.1103/PhysRevB.88.134430} {\bibfield  {journal} {\bibinfo
  {journal} {Phys. Rev. B}\ }\textbf {\bibinfo {volume} {88}},\ \bibinfo
  {pages} {134430} (\bibinfo {year} {2013})}\BibitemShut {NoStop}%
\bibitem [{\citenamefont {Dimitrakis}\ \emph {et~al.}(2016)\citenamefont
  {Dimitrakis}, \citenamefont {Tsongidis},\ and\ \citenamefont
  {Konstandopoulos}}]{Dimitrakis2016-dft-gap}%
  \BibitemOpen
  \bibfield  {author} {\bibinfo {author} {\bibfnamefont {D.~A.}\ \bibnamefont
  {Dimitrakis}}, \bibinfo {author} {\bibfnamefont {N.}~\bibnamefont
  {Tsongidis}},\ and\ \bibinfo {author} {\bibfnamefont {A.}~\bibnamefont
  {Konstandopoulos}},\ }\href {https://doi.org/10.1039/c6cp05073e} {\bibfield
  {journal} {\bibinfo  {journal} {Phys. Chem. Chem. Phys.}\ }\textbf {\bibinfo
  {volume} {18}},\ \bibinfo {pages} {23587} (\bibinfo {year}
  {2016})}\BibitemShut {NoStop}%
\bibitem [{\citenamefont {Taffa}\ \emph {et~al.}(2016)\citenamefont {Taffa},
  \citenamefont {Dillert}, \citenamefont {Ulpe}, \citenamefont {Bauerfeind},
  \citenamefont {Bredow}, \citenamefont {Bahnemann},\ and\ \citenamefont
  {Wark}}]{Paji2004-dft-gap}%
  \BibitemOpen
  \bibfield  {author} {\bibinfo {author} {\bibfnamefont {D.~H.}\ \bibnamefont
  {Taffa}}, \bibinfo {author} {\bibfnamefont {R.}~\bibnamefont {Dillert}},
  \bibinfo {author} {\bibfnamefont {A.~C.}\ \bibnamefont {Ulpe}}, \bibinfo
  {author} {\bibfnamefont {K.~C.~L.}\ \bibnamefont {Bauerfeind}}, \bibinfo
  {author} {\bibfnamefont {T.}~\bibnamefont {Bredow}}, \bibinfo {author}
  {\bibfnamefont {D.~W.}\ \bibnamefont {Bahnemann}},\ and\ \bibinfo {author}
  {\bibfnamefont {M.}~\bibnamefont {Wark}},\ }\href
  {https://doi.org/10.1117/1.JPE.7.012009} {\bibfield  {journal} {\bibinfo
  {journal} {J. Photonics Energy}\ }\textbf {\bibinfo {volume} {7}},\ \bibinfo
  {pages} {1} (\bibinfo {year} {2016})}\BibitemShut {NoStop}%
\bibitem [{\citenamefont {Perdew}\ \emph {et~al.}(1996)\citenamefont {Perdew},
  \citenamefont {Burke},\ and\ \citenamefont {Ernzerhof}}]{Perdew1996-PBE}%
  \BibitemOpen
  \bibfield  {author} {\bibinfo {author} {\bibfnamefont {J.~P.}\ \bibnamefont
  {Perdew}}, \bibinfo {author} {\bibfnamefont {K.}~\bibnamefont {Burke}},\ and\
  \bibinfo {author} {\bibfnamefont {M.}~\bibnamefont {Ernzerhof}},\ }\href
  {https://doi.org/10.1103/PhysRevLett.77.3865} {\bibfield  {journal} {\bibinfo
   {journal} {Phys. Rev. Lett.}\ }\textbf {\bibinfo {volume} {77}},\ \bibinfo
  {pages} {3865} (\bibinfo {year} {1996})}\BibitemShut {NoStop}%
\bibitem [{\citenamefont {Caffrey}\ \emph {et~al.}(2013)\citenamefont
  {Caffrey}, \citenamefont {Fritsch}, \citenamefont {Archer}, \citenamefont
  {Sanvito},\ and\ \citenamefont {Ederer}}]{caffrey2013spin}%
  \BibitemOpen
  \bibfield  {author} {\bibinfo {author} {\bibfnamefont {N.~M.}\ \bibnamefont
  {Caffrey}}, \bibinfo {author} {\bibfnamefont {D.}~\bibnamefont {Fritsch}},
  \bibinfo {author} {\bibfnamefont {T.}~\bibnamefont {Archer}}, \bibinfo
  {author} {\bibfnamefont {S.}~\bibnamefont {Sanvito}},\ and\ \bibinfo {author}
  {\bibfnamefont {C.}~\bibnamefont {Ederer}},\ }\href
  {https://doi.org/10.1103/PhysRevB.87.024419} {\bibfield  {journal} {\bibinfo
  {journal} {Phys. Rev. B}\ }\textbf {\bibinfo {volume} {87}},\ \bibinfo
  {pages} {024419} (\bibinfo {year} {2013})}\BibitemShut {NoStop}%
\bibitem [{\citenamefont {Kresse}\ and\ \citenamefont
  {Furthm\"uller}(1996)}]{kresse199614251}%
  \BibitemOpen
  \bibfield  {author} {\bibinfo {author} {\bibfnamefont {G.}~\bibnamefont
  {Kresse}}\ and\ \bibinfo {author} {\bibfnamefont {J.}~\bibnamefont
  {Furthm\"uller}},\ }\href {https://doi.org/10.1103/PhysRevB.54.11169}
  {\bibfield  {journal} {\bibinfo  {journal} {Phys. Rev. B}\ }\textbf {\bibinfo
  {volume} {54}},\ \bibinfo {pages} {11169} (\bibinfo {year}
  {1996})}\BibitemShut {NoStop}%
\bibitem [{\citenamefont {Kresse}\ and\ \citenamefont
  {Joubert}(1999)}]{kresse1999ultrasoft}%
  \BibitemOpen
  \bibfield  {author} {\bibinfo {author} {\bibfnamefont {G.}~\bibnamefont
  {Kresse}}\ and\ \bibinfo {author} {\bibfnamefont {D.}~\bibnamefont
  {Joubert}},\ }\href {https://doi.org/10.1103/PhysRevB.59.1758} {\bibfield
  {journal} {\bibinfo  {journal} {Phys. Rev. B}\ }\textbf {\bibinfo {volume}
  {59}},\ \bibinfo {pages} {1758} (\bibinfo {year} {1999})}\BibitemShut
  {NoStop}%
\bibitem [{\citenamefont {Giannozzi}\ \emph {et~al.}(2009)\citenamefont
  {Giannozzi}, \citenamefont {Baroni}, \citenamefont {Bonini}, \citenamefont
  {Calandra}, \citenamefont {Car}, \citenamefont {Cavazzoni}, \citenamefont
  {Ceresoli}, \citenamefont {Chiarotti}, \citenamefont {Cococcioni},
  \citenamefont {Dabo}, \citenamefont {Dal~Corso}, \citenamefont
  {de~Gironcoli}, \citenamefont {Fabris}, \citenamefont {Fratesi},
  \citenamefont {Gebauer}, \citenamefont {Gerstmann}, \citenamefont
  {Gougoussis}, \citenamefont {Kokalj}, \citenamefont {Lazzeri}, \citenamefont
  {Martin-Samos}, \citenamefont {Marzari}, \citenamefont {Mauri}, \citenamefont
  {Mazzarello}, \citenamefont {Paolini}, \citenamefont {Pasquarello},
  \citenamefont {Paulatto}, \citenamefont {Sbraccia}, \citenamefont {Scandolo},
  \citenamefont {Sclauzero}, \citenamefont {Seitsonen}, \citenamefont
  {Smogunov}, \citenamefont {Umari},\ and\ \citenamefont
  {Wentzcovitch}}]{Giannozzi2009-QE}%
  \BibitemOpen
  \bibfield  {author} {\bibinfo {author} {\bibfnamefont {P.}~\bibnamefont
  {Giannozzi}}, \bibinfo {author} {\bibfnamefont {S.}~\bibnamefont {Baroni}},
  \bibinfo {author} {\bibfnamefont {N.}~\bibnamefont {Bonini}}, \bibinfo
  {author} {\bibfnamefont {M.}~\bibnamefont {Calandra}}, \bibinfo {author}
  {\bibfnamefont {R.}~\bibnamefont {Car}}, \bibinfo {author} {\bibfnamefont
  {C.}~\bibnamefont {Cavazzoni}}, \bibinfo {author} {\bibfnamefont
  {D.}~\bibnamefont {Ceresoli}}, \bibinfo {author} {\bibfnamefont {G.~L.}\
  \bibnamefont {Chiarotti}}, \bibinfo {author} {\bibfnamefont {M.}~\bibnamefont
  {Cococcioni}}, \bibinfo {author} {\bibfnamefont {I.}~\bibnamefont {Dabo}},
  \bibinfo {author} {\bibfnamefont {A.}~\bibnamefont {Dal~Corso}}, \bibinfo
  {author} {\bibfnamefont {S.}~\bibnamefont {de~Gironcoli}}, \bibinfo {author}
  {\bibfnamefont {S.}~\bibnamefont {Fabris}}, \bibinfo {author} {\bibfnamefont
  {G.}~\bibnamefont {Fratesi}}, \bibinfo {author} {\bibfnamefont
  {R.}~\bibnamefont {Gebauer}}, \bibinfo {author} {\bibfnamefont
  {U.}~\bibnamefont {Gerstmann}}, \bibinfo {author} {\bibfnamefont
  {C.}~\bibnamefont {Gougoussis}}, \bibinfo {author} {\bibfnamefont
  {A.}~\bibnamefont {Kokalj}}, \bibinfo {author} {\bibfnamefont
  {M.}~\bibnamefont {Lazzeri}}, \bibinfo {author} {\bibfnamefont
  {L.}~\bibnamefont {Martin-Samos}}, \bibinfo {author} {\bibfnamefont
  {N.}~\bibnamefont {Marzari}}, \bibinfo {author} {\bibfnamefont
  {F.}~\bibnamefont {Mauri}}, \bibinfo {author} {\bibfnamefont
  {R.}~\bibnamefont {Mazzarello}}, \bibinfo {author} {\bibfnamefont
  {S.}~\bibnamefont {Paolini}}, \bibinfo {author} {\bibfnamefont
  {A.}~\bibnamefont {Pasquarello}}, \bibinfo {author} {\bibfnamefont
  {L.}~\bibnamefont {Paulatto}}, \bibinfo {author} {\bibfnamefont
  {C.}~\bibnamefont {Sbraccia}}, \bibinfo {author} {\bibfnamefont
  {S.}~\bibnamefont {Scandolo}}, \bibinfo {author} {\bibfnamefont
  {G.}~\bibnamefont {Sclauzero}}, \bibinfo {author} {\bibfnamefont {A.~P.}\
  \bibnamefont {Seitsonen}}, \bibinfo {author} {\bibfnamefont {A.}~\bibnamefont
  {Smogunov}}, \bibinfo {author} {\bibfnamefont {P.}~\bibnamefont {Umari}},\
  and\ \bibinfo {author} {\bibfnamefont {R.~M.}\ \bibnamefont {Wentzcovitch}},\
  }\bibfield  {title} {\bibinfo {title} {Quantum espresso: a modular and
  open-source software project for quantum simulations of materials},\ }\href
  {https://doi.org/10.1088/0953-8984/21/39/395502} {\bibfield  {journal}
  {\bibinfo  {journal} {J. Condens. Matter Phys.}\ }\textbf {\bibinfo {volume}
  {21}},\ \bibinfo {pages} {395502} (\bibinfo {year} {2009})}\BibitemShut
  {NoStop}%
\bibitem [{\citenamefont {Pemmaraju}\ \emph {et~al.}(2007)\citenamefont
  {Pemmaraju}, \citenamefont {Archer}, \citenamefont {S\'anchez-Portal},\ and\
  \citenamefont {Sanvito}}]{pemmaraju2007atomic}%
  \BibitemOpen
  \bibfield  {author} {\bibinfo {author} {\bibfnamefont {C.~D.}\ \bibnamefont
  {Pemmaraju}}, \bibinfo {author} {\bibfnamefont {T.}~\bibnamefont {Archer}},
  \bibinfo {author} {\bibfnamefont {D.}~\bibnamefont {S\'anchez-Portal}},\ and\
  \bibinfo {author} {\bibfnamefont {S.}~\bibnamefont {Sanvito}},\ }\href
  {https://doi.org/10.1103/PhysRevB.75.045101} {\bibfield  {journal} {\bibinfo
  {journal} {Phys. Rev. B}\ }\textbf {\bibinfo {volume} {75}},\ \bibinfo
  {pages} {045101} (\bibinfo {year} {2007})}\BibitemShut {NoStop}%
\bibitem [{\citenamefont {Heyd}\ \emph {et~al.}(2003)\citenamefont {Heyd},
  \citenamefont {Scuseria},\ and\ \citenamefont {Ernzerhof}}]{Heyd2003-HSE}%
  \BibitemOpen
  \bibfield  {author} {\bibinfo {author} {\bibfnamefont {J.}~\bibnamefont
  {Heyd}}, \bibinfo {author} {\bibfnamefont {G.~E.}\ \bibnamefont {Scuseria}},\
  and\ \bibinfo {author} {\bibfnamefont {M.}~\bibnamefont {Ernzerhof}},\ }\href
  {https://doi.org/10.1063/1.1564060} {\bibfield  {journal} {\bibinfo
  {journal} {Chem. Phys.}\ }\textbf {\bibinfo {volume} {118}},\ \bibinfo
  {pages} {8207} (\bibinfo {year} {2003})}\BibitemShut {NoStop}%
\bibitem [{\citenamefont {Hou}\ \emph {et~al.}(2010)\citenamefont {Hou},
  \citenamefont {Zhao}, \citenamefont {Liu}, \citenamefont {Yu}, \citenamefont
  {Zhong}, \citenamefont {Qiu}, \citenamefont {Zeng},\ and\ \citenamefont
  {Wen}}]{Hou2010-inversionT}%
  \BibitemOpen
  \bibfield  {author} {\bibinfo {author} {\bibfnamefont {Y.~H.}\ \bibnamefont
  {Hou}}, \bibinfo {author} {\bibfnamefont {Y.~J.}\ \bibnamefont {Zhao}},
  \bibinfo {author} {\bibfnamefont {Z.~W.}\ \bibnamefont {Liu}}, \bibinfo
  {author} {\bibfnamefont {H.~Y.}\ \bibnamefont {Yu}}, \bibinfo {author}
  {\bibfnamefont {X.~C.}\ \bibnamefont {Zhong}}, \bibinfo {author}
  {\bibfnamefont {W.~Q.}\ \bibnamefont {Qiu}}, \bibinfo {author} {\bibfnamefont
  {D.~C.}\ \bibnamefont {Zeng}},\ and\ \bibinfo {author} {\bibfnamefont
  {L.~S.}\ \bibnamefont {Wen}},\ }\href
  {https://doi.org/10.1088/0022-3727/43/44/445003} {\bibfield  {journal}
  {\bibinfo  {journal} {J. Phys. D: Appl. Phys}\ }\textbf {\bibinfo {volume}
  {43}},\ \bibinfo {pages} {445003} (\bibinfo {year} {2010})}\BibitemShut
  {NoStop}%
\bibitem [{\citenamefont {Hedin}(1965)}]{hedin1965new}%
  \BibitemOpen
  \bibfield  {author} {\bibinfo {author} {\bibfnamefont {L.}~\bibnamefont
  {Hedin}},\ }\href {https://doi.org/10.1103/PhysRev.139.A796} {\bibfield
  {journal} {\bibinfo  {journal} {Phys. Rev. B}\ }\textbf {\bibinfo {volume}
  {139}},\ \bibinfo {pages} {A796} (\bibinfo {year} {1965})}\BibitemShut
  {NoStop}%
\bibitem [{\citenamefont {Smart}\ \emph {et~al.}(2019)\citenamefont {Smart},
  \citenamefont {Pham}, \citenamefont {Ping},\ and\ \citenamefont
  {Ogitsu}}]{Smart2019-co3o4-polarons}%
  \BibitemOpen
  \bibfield  {author} {\bibinfo {author} {\bibfnamefont {T.~J.}\ \bibnamefont
  {Smart}}, \bibinfo {author} {\bibfnamefont {T.~A.}\ \bibnamefont {Pham}},
  \bibinfo {author} {\bibfnamefont {Y.}~\bibnamefont {Ping}},\ and\ \bibinfo
  {author} {\bibfnamefont {T.}~\bibnamefont {Ogitsu}},\ }\href
  {https://doi.org/10.1103/PhysRevMaterials.3.102401} {\bibfield  {journal}
  {\bibinfo  {journal} {Phys. Rev. Mater.}\ }\textbf {\bibinfo {volume} {3}},\
  \bibinfo {pages} {102401} (\bibinfo {year} {2019})}\BibitemShut {NoStop}%
\bibitem [{\citenamefont {Singh}\ \emph {et~al.}(2015)\citenamefont {Singh},
  \citenamefont {Kosa}, \citenamefont {Majhi},\ and\ \citenamefont
  {Major}}]{singh2015putting}%
  \BibitemOpen
  \bibfield  {author} {\bibinfo {author} {\bibfnamefont {V.}~\bibnamefont
  {Singh}}, \bibinfo {author} {\bibfnamefont {M.}~\bibnamefont {Kosa}},
  \bibinfo {author} {\bibfnamefont {K.}~\bibnamefont {Majhi}},\ and\ \bibinfo
  {author} {\bibfnamefont {D.~T.}\ \bibnamefont {Major}},\ }\href
  {https://doi.org/10.1021/ct500770m} {\bibfield  {journal} {\bibinfo
  {journal} {J. Chem. Theory Comput.}\ }\textbf {\bibinfo {volume} {11}},\
  \bibinfo {pages} {64} (\bibinfo {year} {2015})}\BibitemShut {NoStop}%
\bibitem [{\citenamefont {Ulpe}\ and\ \citenamefont
  {Bredow}(2020)}]{Ulpe2020-BSE-inversion}%
  \BibitemOpen
  \bibfield  {author} {\bibinfo {author} {\bibfnamefont {A.~C.}\ \bibnamefont
  {Ulpe}}\ and\ \bibinfo {author} {\bibfnamefont {T.}~\bibnamefont {Bredow}},\
  }\href {https://doi.org/10.1002/cphc.201901088} {\bibfield  {journal}
  {\bibinfo  {journal} {ChemPhysChem}\ }\textbf {\bibinfo {volume} {21}},\
  \bibinfo {pages} {546} (\bibinfo {year} {2020})}\BibitemShut {NoStop}%
\bibitem [{\citenamefont {Jiang}\ \emph {et~al.}(2009)\citenamefont {Jiang},
  \citenamefont {Gomez-Abal}, \citenamefont {Rinke},\ and\ \citenamefont
  {Scheffler}}]{Jiang2009-G0W0-LDAU}%
  \BibitemOpen
  \bibfield  {author} {\bibinfo {author} {\bibfnamefont {H.}~\bibnamefont
  {Jiang}}, \bibinfo {author} {\bibfnamefont {R.~I.}\ \bibnamefont
  {Gomez-Abal}}, \bibinfo {author} {\bibfnamefont {P.}~\bibnamefont {Rinke}},\
  and\ \bibinfo {author} {\bibfnamefont {M.}~\bibnamefont {Scheffler}},\ }\href
  {https://doi.org/10.1103/PhysRevLett.102.126403} {\bibfield  {journal}
  {\bibinfo  {journal} {Phys. Rev. Lett.}\ }\textbf {\bibinfo {volume} {102}},\
  \bibinfo {pages} {126403} (\bibinfo {year} {2009})}\BibitemShut {NoStop}%
\bibitem [{\citenamefont {Jiang}\ \emph {et~al.}(2010)\citenamefont {Jiang},
  \citenamefont {Gomez-Abal}, \citenamefont {Rinke},\ and\ \citenamefont
  {Scheffler}}]{Jiang2010-GW-LDAU}%
  \BibitemOpen
  \bibfield  {author} {\bibinfo {author} {\bibfnamefont {H.}~\bibnamefont
  {Jiang}}, \bibinfo {author} {\bibfnamefont {R.~I.}\ \bibnamefont
  {Gomez-Abal}}, \bibinfo {author} {\bibfnamefont {P.}~\bibnamefont {Rinke}},\
  and\ \bibinfo {author} {\bibfnamefont {M.}~\bibnamefont {Scheffler}},\ }\href
  {https://doi.org/10.1103/PhysRevB.82.045108} {\bibfield  {journal} {\bibinfo
  {journal} {Phys. Rev. B}\ }\textbf {\bibinfo {volume} {82}},\ \bibinfo
  {pages} {045108} (\bibinfo {year} {2010})}\BibitemShut {NoStop}%
\bibitem [{\citenamefont {Jiang}\ \emph {et~al.}(2012)\citenamefont {Jiang},
  \citenamefont {Lu}, \citenamefont {Long},\ and\ \citenamefont
  {Chen}}]{jiang2012ab}%
  \BibitemOpen
  \bibfield  {author} {\bibinfo {author} {\bibfnamefont {S.}~\bibnamefont
  {Jiang}}, \bibinfo {author} {\bibfnamefont {T.}~\bibnamefont {Lu}}, \bibinfo
  {author} {\bibfnamefont {Y.}~\bibnamefont {Long}},\ and\ \bibinfo {author}
  {\bibfnamefont {J.}~\bibnamefont {Chen}},\ }\href
  {https://doi.org/10.1063/1.3686727} {\bibfield  {journal} {\bibinfo
  {journal} {J. Appl. Phys.}\ }\textbf {\bibinfo {volume} {111}},\ \bibinfo
  {pages} {043516} (\bibinfo {year} {2012})}\BibitemShut {NoStop}%
\bibitem [{\citenamefont {Lany}(2013)}]{lany2013band}%
  \BibitemOpen
  \bibfield  {author} {\bibinfo {author} {\bibfnamefont {S.}~\bibnamefont
  {Lany}},\ }\href {https://doi.org/10.1103/PhysRevB.87.085112} {\bibfield
  {journal} {\bibinfo  {journal} {Phys. Rev. B}\ }\textbf {\bibinfo {volume}
  {87}},\ \bibinfo {pages} {085112} (\bibinfo {year} {2013})}\BibitemShut
  {NoStop}%
\bibitem [{\citenamefont {Piccinin}(2019)}]{piccinin2019band}%
  \BibitemOpen
  \bibfield  {author} {\bibinfo {author} {\bibfnamefont {S.}~\bibnamefont
  {Piccinin}},\ }\href {https://doi.org/10.1039/c8cp07132b} {\bibfield
  {journal} {\bibinfo  {journal} {Phys. Chem. Chem. Phys.}\ }\textbf {\bibinfo
  {volume} {21}},\ \bibinfo {pages} {2957} (\bibinfo {year}
  {2019})}\BibitemShut {NoStop}%
\bibitem [{\citenamefont {Rohlfing}\ and\ \citenamefont
  {Louie}(1998)}]{Rohlfing1998-BSE}%
  \BibitemOpen
  \bibfield  {author} {\bibinfo {author} {\bibfnamefont {M.}~\bibnamefont
  {Rohlfing}}\ and\ \bibinfo {author} {\bibfnamefont {S.~G.}\ \bibnamefont
  {Louie}},\ }\bibfield  {title} {\bibinfo {title} {Electron-hole excitations
  in semiconductors and insulators},\ }\href
  {https://doi.org/10.1103/PhysRevLett.81.2312} {\bibfield  {journal} {\bibinfo
   {journal} {Phys. Rev. Lett.}\ }\textbf {\bibinfo {volume} {81}},\ \bibinfo
  {pages} {2312} (\bibinfo {year} {1998})}\BibitemShut {NoStop}%
\bibitem [{\citenamefont {Radha}\ \emph {et~al.}(2021)\citenamefont {Radha},
  \citenamefont {Lambrecht}, \citenamefont {Cunningham}, \citenamefont
  {Gr\"uning}, \citenamefont {Pashov},\ and\ \citenamefont {van
  Schilfgaarde}}]{Radha2021-LiCoO2}%
  \BibitemOpen
  \bibfield  {author} {\bibinfo {author} {\bibfnamefont {S.~K.}\ \bibnamefont
  {Radha}}, \bibinfo {author} {\bibfnamefont {W.~R.~L.}\ \bibnamefont
  {Lambrecht}}, \bibinfo {author} {\bibfnamefont {B.}~\bibnamefont
  {Cunningham}}, \bibinfo {author} {\bibfnamefont {M.}~\bibnamefont
  {Gr\"uning}}, \bibinfo {author} {\bibfnamefont {D.}~\bibnamefont {Pashov}},\
  and\ \bibinfo {author} {\bibfnamefont {M.}~\bibnamefont {van Schilfgaarde}},\
  }\href {https://doi.org/10.1103/PhysRevB.104.115120} {\bibfield  {journal}
  {\bibinfo  {journal} {Phys. Rev. B}\ }\textbf {\bibinfo {volume} {104}},\
  \bibinfo {pages} {115120} (\bibinfo {year} {2021})}\BibitemShut {NoStop}%
\bibitem [{\citenamefont {Sponza}\ \emph {et~al.}(2013)\citenamefont {Sponza},
  \citenamefont {V\'eniard}, \citenamefont {Sottile}, \citenamefont
  {Giorgetti},\ and\ \citenamefont {Reining}}]{Sponza-2013}%
  \BibitemOpen
  \bibfield  {author} {\bibinfo {author} {\bibfnamefont {L.}~\bibnamefont
  {Sponza}}, \bibinfo {author} {\bibfnamefont {V.}~\bibnamefont {V\'eniard}},
  \bibinfo {author} {\bibfnamefont {F.}~\bibnamefont {Sottile}}, \bibinfo
  {author} {\bibfnamefont {C.}~\bibnamefont {Giorgetti}},\ and\ \bibinfo
  {author} {\bibfnamefont {L.}~\bibnamefont {Reining}},\ }\href
  {https://doi.org/10.1103/PhysRevB.87.235102} {\bibfield  {journal} {\bibinfo
  {journal} {Phys. Rev. B}\ }\textbf {\bibinfo {volume} {87}},\ \bibinfo
  {pages} {235102} (\bibinfo {year} {2013})}\BibitemShut {NoStop}%
\bibitem [{\citenamefont {Begum}\ \emph {et~al.}(2019)\citenamefont {Begum},
  \citenamefont {Gruner},\ and\ \citenamefont {Pentcheva}}]{Begum2019-p1}%
  \BibitemOpen
  \bibfield  {author} {\bibinfo {author} {\bibfnamefont {V.}~\bibnamefont
  {Begum}}, \bibinfo {author} {\bibfnamefont {M.~E.}\ \bibnamefont {Gruner}},\
  and\ \bibinfo {author} {\bibfnamefont {R.}~\bibnamefont {Pentcheva}},\ }\href
  {https://doi.org/10.1103/PhysRevMaterials.3.065004} {\bibfield  {journal}
  {\bibinfo  {journal} {Phys. Rev. Mater.}\ }\textbf {\bibinfo {volume} {3}},\
  \bibinfo {pages} {065004} (\bibinfo {year} {2019})}\BibitemShut {NoStop}%
\bibitem [{\citenamefont {Wang}\ \emph {et~al.}(2004)\citenamefont {Wang},
  \citenamefont {Rohlfing}, \citenamefont {Kr{\"u}ger},\ and\ \citenamefont
  {Pollmann}}]{Wang2004}%
  \BibitemOpen
  \bibfield  {author} {\bibinfo {author} {\bibfnamefont {N.-P.}\ \bibnamefont
  {Wang}}, \bibinfo {author} {\bibfnamefont {M.}~\bibnamefont {Rohlfing}},
  \bibinfo {author} {\bibfnamefont {P.}~\bibnamefont {Kr{\"u}ger}},\ and\
  \bibinfo {author} {\bibfnamefont {J.}~\bibnamefont {Pollmann}},\ }\href
  {https://doi.org/10.1007/s00339-003-2305-3} {\bibfield  {journal} {\bibinfo
  {journal} {Appl. Phys. A}\ }\textbf {\bibinfo {volume} {78}},\ \bibinfo
  {pages} {213} (\bibinfo {year} {2004})}\BibitemShut {NoStop}%
\bibitem [{\citenamefont {Begum}\ \emph {et~al.}(2021)\citenamefont {Begum},
  \citenamefont {Gruner}, \citenamefont {Vorwerk}, \citenamefont {Draxl},\ and\
  \citenamefont {Pentcheva}}]{Begum2021-p2}%
  \BibitemOpen
  \bibfield  {author} {\bibinfo {author} {\bibfnamefont {V.}~\bibnamefont
  {Begum}}, \bibinfo {author} {\bibfnamefont {M.~E.}\ \bibnamefont {Gruner}},
  \bibinfo {author} {\bibfnamefont {C.}~\bibnamefont {Vorwerk}}, \bibinfo
  {author} {\bibfnamefont {C.}~\bibnamefont {Draxl}},\ and\ \bibinfo {author}
  {\bibfnamefont {R.}~\bibnamefont {Pentcheva}},\ }\href
  {https://doi.org/10.1103/PhysRevB.103.195128} {\bibfield  {journal} {\bibinfo
   {journal} {Phys. Rev. B}\ }\textbf {\bibinfo {volume} {103}},\ \bibinfo
  {pages} {195128} (\bibinfo {year} {2021})}\BibitemShut {NoStop}%
\bibitem [{\citenamefont {Sun}\ \emph {et~al.}(2015)\citenamefont {Sun},
  \citenamefont {Ruzsinszky},\ and\ \citenamefont {Perdew}}]{Sun2015-SCAN}%
  \BibitemOpen
  \bibfield  {author} {\bibinfo {author} {\bibfnamefont {J.}~\bibnamefont
  {Sun}}, \bibinfo {author} {\bibfnamefont {A.}~\bibnamefont {Ruzsinszky}},\
  and\ \bibinfo {author} {\bibfnamefont {J.~P.}\ \bibnamefont {Perdew}},\
  }\href {https://doi.org/10.1103/PhysRevLett.115.036402} {\bibfield  {journal}
  {\bibinfo  {journal} {Phys. Rev. Lett.}\ }\textbf {\bibinfo {volume} {115}},\
  \bibinfo {pages} {036402} (\bibinfo {year} {2015})}\BibitemShut {NoStop}%
\bibitem [{\citenamefont {Krukau}\ \emph {et~al.}(2006)\citenamefont {Krukau},
  \citenamefont {Vydrov}, \citenamefont {Izmaylov},\ and\ \citenamefont
  {Scuseria}}]{Krukau2006-HSE06}%
  \BibitemOpen
  \bibfield  {author} {\bibinfo {author} {\bibfnamefont {A.~V.}\ \bibnamefont
  {Krukau}}, \bibinfo {author} {\bibfnamefont {O.~A.}\ \bibnamefont {Vydrov}},
  \bibinfo {author} {\bibfnamefont {A.~F.}\ \bibnamefont {Izmaylov}},\ and\
  \bibinfo {author} {\bibfnamefont {G.~E.}\ \bibnamefont {Scuseria}},\ }\href
  {https://doi.org/10.1063/1.2404663} {\bibfield  {journal} {\bibinfo
  {journal} {J. Chem. Phys.}\ }\textbf {\bibinfo {volume} {125}},\ \bibinfo
  {pages} {224106} (\bibinfo {year} {2006})}\BibitemShut {NoStop}%
\bibitem [{\citenamefont {Bokdam}\ \emph {et~al.}(2016)\citenamefont {Bokdam},
  \citenamefont {Sander}, \citenamefont {Stroppa}, \citenamefont {Picozzi},
  \citenamefont {Sarma}, \citenamefont {Franchini},\ and\ \citenamefont
  {Kresse}}]{Bokdam2016-MBSE}%
  \BibitemOpen
  \bibfield  {author} {\bibinfo {author} {\bibfnamefont {M.}~\bibnamefont
  {Bokdam}}, \bibinfo {author} {\bibfnamefont {T.}~\bibnamefont {Sander}},
  \bibinfo {author} {\bibfnamefont {A.}~\bibnamefont {Stroppa}}, \bibinfo
  {author} {\bibfnamefont {S.}~\bibnamefont {Picozzi}}, \bibinfo {author}
  {\bibfnamefont {D.~D.}\ \bibnamefont {Sarma}}, \bibinfo {author}
  {\bibfnamefont {C.}~\bibnamefont {Franchini}},\ and\ \bibinfo {author}
  {\bibfnamefont {G.}~\bibnamefont {Kresse}},\ }\href
  {https://doi.org/10.1038/srep28618} {\bibfield  {journal} {\bibinfo
  {journal} {Sci. Reports}\ }\textbf {\bibinfo {volume} {6}},\ \bibinfo {pages}
  {28618} (\bibinfo {year} {2016})}\BibitemShut {NoStop}%
\bibitem [{\citenamefont {Fuchs}\ \emph {et~al.}(2008)\citenamefont {Fuchs},
  \citenamefont {R\"odl}, \citenamefont {Schleife},\ and\ \citenamefont
  {Bechstedt}}]{Fuchs2008-mBSE}%
  \BibitemOpen
  \bibfield  {author} {\bibinfo {author} {\bibfnamefont {F.}~\bibnamefont
  {Fuchs}}, \bibinfo {author} {\bibfnamefont {C.}~\bibnamefont {R\"odl}},
  \bibinfo {author} {\bibfnamefont {A.}~\bibnamefont {Schleife}},\ and\
  \bibinfo {author} {\bibfnamefont {F.}~\bibnamefont {Bechstedt}},\ }\href
  {https://doi.org/10.1103/PhysRevB.78.085103} {\bibfield  {journal} {\bibinfo
  {journal} {Phys. Rev. B}\ }\textbf {\bibinfo {volume} {78}},\ \bibinfo
  {pages} {085103} (\bibinfo {year} {2008})}\BibitemShut {NoStop}%
\bibitem [{\citenamefont {Liu}\ \emph {et~al.}(2018)\citenamefont {Liu},
  \citenamefont {Kim}, \citenamefont {Chen}, \citenamefont {Sarma},
  \citenamefont {Kresse},\ and\ \citenamefont {Franchini}}]{Liu2018-mBSE}%
  \BibitemOpen
  \bibfield  {author} {\bibinfo {author} {\bibfnamefont {P.}~\bibnamefont
  {Liu}}, \bibinfo {author} {\bibfnamefont {B.}~\bibnamefont {Kim}}, \bibinfo
  {author} {\bibfnamefont {X.-Q.}\ \bibnamefont {Chen}}, \bibinfo {author}
  {\bibfnamefont {D.~D.}\ \bibnamefont {Sarma}}, \bibinfo {author}
  {\bibfnamefont {G.}~\bibnamefont {Kresse}},\ and\ \bibinfo {author}
  {\bibfnamefont {C.}~\bibnamefont {Franchini}},\ }\href
  {https://doi.org/10.1103/PhysRevMaterials.2.075003} {\bibfield  {journal}
  {\bibinfo  {journal} {Phys. Rev. Mater.}\ }\textbf {\bibinfo {volume} {2}},\
  \bibinfo {pages} {075003} (\bibinfo {year} {2018})}\BibitemShut {NoStop}%
\bibitem [{\citenamefont {Bechstedt}\ \emph {et~al.}(1992)\citenamefont
  {Bechstedt}, \citenamefont {Del~Sole}, \citenamefont {Cappellini},\ and\
  \citenamefont {Reining}}]{Bechstedt1992-mBSE}%
  \BibitemOpen
  \bibfield  {author} {\bibinfo {author} {\bibfnamefont {F.}~\bibnamefont
  {Bechstedt}}, \bibinfo {author} {\bibfnamefont {R.}~\bibnamefont {Del~Sole}},
  \bibinfo {author} {\bibfnamefont {G.}~\bibnamefont {Cappellini}},\ and\
  \bibinfo {author} {\bibfnamefont {L.}~\bibnamefont {Reining}},\ }\href
  {https://www.sciencedirect.com/science/article/pii/003810989290476P}
  {\bibfield  {journal} {\bibinfo  {journal} {Solid State Commun.}\ }\textbf
  {\bibinfo {volume} {84}},\ \bibinfo {pages} {765} (\bibinfo {year}
  {1992})}\BibitemShut {NoStop}%
\bibitem [{\citenamefont {Varrassi}\ \emph {et~al.}(2021)\citenamefont
  {Varrassi}, \citenamefont {Liu}, \citenamefont {Yavas}, \citenamefont
  {Bokdam}, \citenamefont {Kresse},\ and\ \citenamefont
  {Franchini}}]{Varrassi2021-mBSE}%
  \BibitemOpen
  \bibfield  {author} {\bibinfo {author} {\bibfnamefont {L.}~\bibnamefont
  {Varrassi}}, \bibinfo {author} {\bibfnamefont {P.}~\bibnamefont {Liu}},
  \bibinfo {author} {\bibfnamefont {Z.~E.}\ \bibnamefont {Yavas}}, \bibinfo
  {author} {\bibfnamefont {M.}~\bibnamefont {Bokdam}}, \bibinfo {author}
  {\bibfnamefont {G.}~\bibnamefont {Kresse}},\ and\ \bibinfo {author}
  {\bibfnamefont {C.}~\bibnamefont {Franchini}},\ }\href
  {https://doi.org/10.1103/PhysRevMaterials.5.074601} {\bibfield  {journal}
  {\bibinfo  {journal} {Phys. Rev. Mater.}\ }\textbf {\bibinfo {volume} {5}},\
  \bibinfo {pages} {074601} (\bibinfo {year} {2021})}\BibitemShut {NoStop}%
\bibitem [{\citenamefont {Singh}\ \emph
  {et~al.}(2020{\natexlab{b}})\citenamefont {Singh}, \citenamefont {Park},
  \citenamefont {Singh}, \citenamefont {Kim}, \citenamefont {Lim},
  \citenamefont {Kumar}, \citenamefont {Kim}, \citenamefont {Lee},\ and\
  \citenamefont {Chae}}]{Singh2020-LC}%
  \BibitemOpen
  \bibfield  {author} {\bibinfo {author} {\bibfnamefont {J.~P.}\ \bibnamefont
  {Singh}}, \bibinfo {author} {\bibfnamefont {J.~Y.}\ \bibnamefont {Park}},
  \bibinfo {author} {\bibfnamefont {V.}~\bibnamefont {Singh}}, \bibinfo
  {author} {\bibfnamefont {S.~H.}\ \bibnamefont {Kim}}, \bibinfo {author}
  {\bibfnamefont {W.~C.}\ \bibnamefont {Lim}}, \bibinfo {author} {\bibfnamefont
  {H.}~\bibnamefont {Kumar}}, \bibinfo {author} {\bibfnamefont {Y.~H.}\
  \bibnamefont {Kim}}, \bibinfo {author} {\bibfnamefont {S.}~\bibnamefont
  {Lee}},\ and\ \bibinfo {author} {\bibfnamefont {K.~H.}\ \bibnamefont
  {Chae}},\ }\href {https://doi.org/10.1039/d0ra01653e} {\bibfield  {journal}
  {\bibinfo  {journal} {RSC Adv.}\ }\textbf {\bibinfo {volume} {10}},\ \bibinfo
  {pages} {21259} (\bibinfo {year} {2020}{\natexlab{b}})}\BibitemShut {NoStop}%
\bibitem [{\citenamefont {Martens}\ \emph {et~al.}(1985)\citenamefont
  {Martens}, \citenamefont {Peeters}, \citenamefont {van Noort},\ and\
  \citenamefont {Erman}}]{martens1982-DF-CFO}%
  \BibitemOpen
  \bibfield  {author} {\bibinfo {author} {\bibfnamefont {J.~W.~D.}\
  \bibnamefont {Martens}}, \bibinfo {author} {\bibfnamefont {W.~L.}\
  \bibnamefont {Peeters}}, \bibinfo {author} {\bibfnamefont {H.~M.}\
  \bibnamefont {van Noort}},\ and\ \bibinfo {author} {\bibfnamefont
  {M.}~\bibnamefont {Erman}},\ }\href
  {https://www.sciencedirect.com/science/article/pii/0022369785901040}
  {\bibfield  {journal} {\bibinfo  {journal} {J. Phys. Chem. Solids}\ }\textbf
  {\bibinfo {volume} {46}},\ \bibinfo {pages} {411} (\bibinfo {year}
  {1985})}\BibitemShut {NoStop}%
\bibitem [{\citenamefont {Li}\ \emph {et~al.}(1991)\citenamefont {Li},
  \citenamefont {Fisher}, \citenamefont {Liu},\ and\ \citenamefont
  {Nevitt}}]{li1991single}%
  \BibitemOpen
  \bibfield  {author} {\bibinfo {author} {\bibfnamefont {Z.}~\bibnamefont
  {Li}}, \bibinfo {author} {\bibfnamefont {E.~S.}\ \bibnamefont {Fisher}},
  \bibinfo {author} {\bibfnamefont {J.~Z.}\ \bibnamefont {Liu}},\ and\ \bibinfo
  {author} {\bibfnamefont {M.~V.}\ \bibnamefont {Nevitt}},\ }\href
  {https://doi.org/10.1007/BF02387728} {\bibfield  {journal} {\bibinfo
  {journal} {J. Mater. Sci.}\ }\textbf {\bibinfo {volume} {26}},\ \bibinfo
  {pages} {2621} (\bibinfo {year} {1991})}\BibitemShut {NoStop}%
\bibitem [{\citenamefont {Bl\"ochl}(1994)}]{blochl1994projector}%
  \BibitemOpen
  \bibfield  {author} {\bibinfo {author} {\bibfnamefont {P.~E.}\ \bibnamefont
  {Bl\"ochl}},\ }\href {https://doi.org/10.1103/PhysRevB.50.17953} {\bibfield
  {journal} {\bibinfo  {journal} {Phys. Rev. B}\ }\textbf {\bibinfo {volume}
  {50}},\ \bibinfo {pages} {17953} (\bibinfo {year} {1994})}\BibitemShut
  {NoStop}%
\bibitem [{\citenamefont {Dudarev}\ \emph {et~al.}(1998)\citenamefont
  {Dudarev}, \citenamefont {Botton}, \citenamefont {Savrasov}, \citenamefont
  {Humphreys},\ and\ \citenamefont {Sutton}}]{Dudarev-HubbardU}%
  \BibitemOpen
  \bibfield  {author} {\bibinfo {author} {\bibfnamefont {S.~L.}\ \bibnamefont
  {Dudarev}}, \bibinfo {author} {\bibfnamefont {G.~A.}\ \bibnamefont {Botton}},
  \bibinfo {author} {\bibfnamefont {S.~Y.}\ \bibnamefont {Savrasov}}, \bibinfo
  {author} {\bibfnamefont {C.~J.}\ \bibnamefont {Humphreys}},\ and\ \bibinfo
  {author} {\bibfnamefont {A.~P.}\ \bibnamefont {Sutton}},\ }\href
  {https://doi.org/10.1103/PhysRevB.57.1505} {\bibfield  {journal} {\bibinfo
  {journal} {Phys. Rev. B}\ }\textbf {\bibinfo {volume} {57}},\ \bibinfo
  {pages} {1505} (\bibinfo {year} {1998})}\BibitemShut {NoStop}%
\bibitem [{\citenamefont {Mostofi}\ \emph {et~al.}(2008)\citenamefont
  {Mostofi}, \citenamefont {Yates}, \citenamefont {Lee}, \citenamefont {Souza},
  \citenamefont {Vanderbilt},\ and\ \citenamefont
  {Marzari}}]{mostofi2008wannier90}%
  \BibitemOpen
  \bibfield  {author} {\bibinfo {author} {\bibfnamefont {A.~A.}\ \bibnamefont
  {Mostofi}}, \bibinfo {author} {\bibfnamefont {J.~R.}\ \bibnamefont {Yates}},
  \bibinfo {author} {\bibfnamefont {Y.-S.}\ \bibnamefont {Lee}}, \bibinfo
  {author} {\bibfnamefont {I.}~\bibnamefont {Souza}}, \bibinfo {author}
  {\bibfnamefont {D.}~\bibnamefont {Vanderbilt}},\ and\ \bibinfo {author}
  {\bibfnamefont {N.}~\bibnamefont {Marzari}},\ }\href
  {https://www.sciencedirect.com/science/article/pii/S0010465507004936}
  {\bibfield  {journal} {\bibinfo  {journal} {Comput. Phys. Commun.}\ }\textbf
  {\bibinfo {volume} {178}},\ \bibinfo {pages} {685} (\bibinfo {year}
  {2008})}\BibitemShut {NoStop}%
\bibitem [{\citenamefont {Stokes}\ and\ \citenamefont
  {Hatch}(2005)}]{Stokes2005-FINDSYM}%
  \BibitemOpen
  \bibfield  {author} {\bibinfo {author} {\bibfnamefont {H.~T.}\ \bibnamefont
  {Stokes}}\ and\ \bibinfo {author} {\bibfnamefont {D.~M.}\ \bibnamefont
  {Hatch}},\ }\href {https://doi.org/10.1107/S0021889804031528} {\bibfield
  {journal} {\bibinfo  {journal} {J. Appl. Crystallogr.}\ }\textbf {\bibinfo
  {volume} {38}},\ \bibinfo {pages} {237} (\bibinfo {year} {2005})}\BibitemShut
  {NoStop}%
\bibitem [{\citenamefont {Dubeck{\'y}}\ \emph {et~al.}(2023)\citenamefont
  {Dubeck{\'y}}, \citenamefont {Min{\'a}rik},\ and\ \citenamefont
  {Karlick{\'y}}}]{Dubecky2023}%
  \BibitemOpen
  \bibfield  {author} {\bibinfo {author} {\bibfnamefont {M.}~\bibnamefont
  {Dubeck{\'y}}}, \bibinfo {author} {\bibfnamefont {S.}~\bibnamefont
  {Min{\'a}rik}},\ and\ \bibinfo {author} {\bibfnamefont {F.}~\bibnamefont
  {Karlick{\'y}}},\ }\href {https://doi.org/10.1063/5.0140315} {\bibfield
  {journal} {\bibinfo  {journal} {J. Chem. Phys.}\ }\textbf {\bibinfo {volume}
  {158}},\ \bibinfo {pages} {054703} (\bibinfo {year} {2023})}\BibitemShut
  {NoStop}%
\bibitem [{\citenamefont {Hanke}\ and\ \citenamefont
  {Sham}(1980)}]{Hanke1980-BSE}%
  \BibitemOpen
  \bibfield  {author} {\bibinfo {author} {\bibfnamefont {W.}~\bibnamefont
  {Hanke}}\ and\ \bibinfo {author} {\bibfnamefont {L.~J.}\ \bibnamefont
  {Sham}},\ }\href {https://doi.org/10.1103/PhysRevB.21.4656} {\bibfield
  {journal} {\bibinfo  {journal} {Phys. Rev. B}\ }\textbf {\bibinfo {volume}
  {21}},\ \bibinfo {pages} {4656} (\bibinfo {year} {1980})}\BibitemShut
  {NoStop}%
\bibitem [{\citenamefont {Shahbahrami}\ \emph {et~al.}(2022)\citenamefont
  {Shahbahrami}, \citenamefont {Rabiee}, \citenamefont {Shidpoor},\ and\
  \citenamefont {Salimi-Kenari}}]{shahbahrami2022exploring}%
  \BibitemOpen
  \bibfield  {author} {\bibinfo {author} {\bibfnamefont {B.}~\bibnamefont
  {Shahbahrami}}, \bibinfo {author} {\bibfnamefont {S.~M.}\ \bibnamefont
  {Rabiee}}, \bibinfo {author} {\bibfnamefont {R.}~\bibnamefont {Shidpoor}},\
  and\ \bibinfo {author} {\bibfnamefont {H.}~\bibnamefont {Salimi-Kenari}},\
  }\href {https://doi.org/10.1007/s11664-022-09512-y} {\bibfield  {journal}
  {\bibinfo  {journal} {J. Electron. Mater.}\ }\textbf {\bibinfo {volume}
  {51}},\ \bibinfo {pages} {2552} (\bibinfo {year} {2022})}\BibitemShut
  {NoStop}%
\bibitem [{\citenamefont {Zviagin}\ \emph {et~al.}(2016)\citenamefont
  {Zviagin}, \citenamefont {Richter}, \citenamefont {B{\"o}ntgen},
  \citenamefont {Lorenz}, \citenamefont {Ziese}, \citenamefont {Zahn},
  \citenamefont {Salvan}, \citenamefont {Grundmann},\ and\ \citenamefont
  {Schmidt-Grund}}]{Zviagin2016-DF-both}%
  \BibitemOpen
  \bibfield  {author} {\bibinfo {author} {\bibfnamefont {V.}~\bibnamefont
  {Zviagin}}, \bibinfo {author} {\bibfnamefont {P.}~\bibnamefont {Richter}},
  \bibinfo {author} {\bibfnamefont {T.}~\bibnamefont {B{\"o}ntgen}}, \bibinfo
  {author} {\bibfnamefont {M.}~\bibnamefont {Lorenz}}, \bibinfo {author}
  {\bibfnamefont {M.}~\bibnamefont {Ziese}}, \bibinfo {author} {\bibfnamefont
  {D.~R.~T.}\ \bibnamefont {Zahn}}, \bibinfo {author} {\bibfnamefont
  {G.}~\bibnamefont {Salvan}}, \bibinfo {author} {\bibfnamefont
  {M.}~\bibnamefont {Grundmann}},\ and\ \bibinfo {author} {\bibfnamefont
  {R.}~\bibnamefont {Schmidt-Grund}},\ }\href
  {https://doi.org/10.1002/pssb.201552361} {\bibfield  {journal} {\bibinfo
  {journal} {Phys. Status Solidi B}\ }\textbf {\bibinfo {volume} {253}},\
  \bibinfo {pages} {429} (\bibinfo {year} {2016})}\BibitemShut {NoStop}%
\bibitem [{\citenamefont {Baldini}\ \emph {et~al.}(2017)\citenamefont
  {Baldini}, \citenamefont {Chiodo}, \citenamefont {Dominguez}, \citenamefont
  {Palummo}, \citenamefont {Moser}, \citenamefont {Yazdi-Rizi}, \citenamefont
  {Aub{\"o}ck}, \citenamefont {Mallett}, \citenamefont {Berger}, \citenamefont
  {Magrez}, \citenamefont {Bernhard}, \citenamefont {Grioni}, \citenamefont
  {Rubio},\ and\ \citenamefont {Chergui}}]{Baldini2017-BE-ex}%
  \BibitemOpen
  \bibfield  {author} {\bibinfo {author} {\bibfnamefont {E.}~\bibnamefont
  {Baldini}}, \bibinfo {author} {\bibfnamefont {L.}~\bibnamefont {Chiodo}},
  \bibinfo {author} {\bibfnamefont {A.}~\bibnamefont {Dominguez}}, \bibinfo
  {author} {\bibfnamefont {M.}~\bibnamefont {Palummo}}, \bibinfo {author}
  {\bibfnamefont {S.}~\bibnamefont {Moser}}, \bibinfo {author} {\bibfnamefont
  {M.}~\bibnamefont {Yazdi-Rizi}}, \bibinfo {author} {\bibfnamefont
  {G.}~\bibnamefont {Aub{\"o}ck}}, \bibinfo {author} {\bibfnamefont {B.~P.~P.}\
  \bibnamefont {Mallett}}, \bibinfo {author} {\bibfnamefont {H.}~\bibnamefont
  {Berger}}, \bibinfo {author} {\bibfnamefont {A.}~\bibnamefont {Magrez}},
  \bibinfo {author} {\bibfnamefont {C.}~\bibnamefont {Bernhard}}, \bibinfo
  {author} {\bibfnamefont {M.}~\bibnamefont {Grioni}}, \bibinfo {author}
  {\bibfnamefont {A.}~\bibnamefont {Rubio}},\ and\ \bibinfo {author}
  {\bibfnamefont {M.}~\bibnamefont {Chergui}},\ }\href
  {https://doi.org/10.1038/s41467-017-00016-6} {\bibfield  {journal} {\bibinfo
  {journal} {Nat. Commun.}\ }\textbf {\bibinfo {volume} {8}},\ \bibinfo {pages}
  {13} (\bibinfo {year} {2017})}\BibitemShut {NoStop}%
\bibitem [{\citenamefont {Snir}\ and\ \citenamefont
  {Toroker}(2020)}]{snir2020operando}%
  \BibitemOpen
  \bibfield  {author} {\bibinfo {author} {\bibfnamefont {N.}~\bibnamefont
  {Snir}}\ and\ \bibinfo {author} {\bibfnamefont {M.~C.}\ \bibnamefont
  {Toroker}},\ }\href {https://doi.org/10.1021/acs.jctc.9b00595} {\bibfield
  {journal} {\bibinfo  {journal} {J. Chem. Theory Comput.}\ }\textbf {\bibinfo
  {volume} {16}},\ \bibinfo {pages} {4857} (\bibinfo {year}
  {2020})}\BibitemShut {NoStop}%
\bibitem [{\citenamefont {Wu}\ \emph {et~al.}(2021)\citenamefont {Wu},
  \citenamefont {Z\"ollner}, \citenamefont {Esser}, \citenamefont {Begum},
  \citenamefont {Prinz}, \citenamefont {Lorke}, \citenamefont {Gegenwart},\
  and\ \citenamefont {Pentcheva}}]{wu2021electronic}%
  \BibitemOpen
  \bibfield  {author} {\bibinfo {author} {\bibfnamefont {J.}~\bibnamefont
  {Wu}}, \bibinfo {author} {\bibfnamefont {M.}~\bibnamefont {Z\"ollner}},
  \bibinfo {author} {\bibfnamefont {S.}~\bibnamefont {Esser}}, \bibinfo
  {author} {\bibfnamefont {V.}~\bibnamefont {Begum}}, \bibinfo {author}
  {\bibfnamefont {G.}~\bibnamefont {Prinz}}, \bibinfo {author} {\bibfnamefont
  {A.}~\bibnamefont {Lorke}}, \bibinfo {author} {\bibfnamefont
  {P.}~\bibnamefont {Gegenwart}},\ and\ \bibinfo {author} {\bibfnamefont
  {R.}~\bibnamefont {Pentcheva}},\ }\href
  {https://doi.org/10.1103/PhysRevB.104.205126} {\bibfield  {journal} {\bibinfo
   {journal} {Phys. Rev. B}\ }\textbf {\bibinfo {volume} {104}},\ \bibinfo
  {pages} {205126} (\bibinfo {year} {2021})}\BibitemShut {NoStop}%
\bibitem [{\citenamefont {Venturini}\ \emph
  {et~al.}(2019{\natexlab{b}})\citenamefont {Venturini}, \citenamefont
  {Tonelli}, \citenamefont {Wermuth}, \citenamefont {Zampiva}, \citenamefont
  {Arcaro}, \citenamefont {Da~Cas~Viegas},\ and\ \citenamefont
  {Bergmann}}]{Venturini2019-INV-CFO}%
  \BibitemOpen
  \bibfield  {author} {\bibinfo {author} {\bibfnamefont {J.}~\bibnamefont
  {Venturini}}, \bibinfo {author} {\bibfnamefont {A.~M.}\ \bibnamefont
  {Tonelli}}, \bibinfo {author} {\bibfnamefont {T.~B.}\ \bibnamefont
  {Wermuth}}, \bibinfo {author} {\bibfnamefont {R.~Y.~S.}\ \bibnamefont
  {Zampiva}}, \bibinfo {author} {\bibfnamefont {S.}~\bibnamefont {Arcaro}},
  \bibinfo {author} {\bibfnamefont {A.}~\bibnamefont {Da~Cas~Viegas}},\ and\
  \bibinfo {author} {\bibfnamefont {C.~P.}\ \bibnamefont {Bergmann}},\ }\href
  {https://www.sciencedirect.com/science/article/pii/S0304885318333432}
  {\bibfield  {journal} {\bibinfo  {journal} {J. Magn. Magn. Mater.}\ }\textbf
  {\bibinfo {volume} {482}},\ \bibinfo {pages} {1} (\bibinfo {year}
  {2019}{\natexlab{b}})}\BibitemShut {NoStop}%
\bibitem [{\citenamefont {Lohaus}\ \emph {et~al.}(2018)\citenamefont {Lohaus},
  \citenamefont {Klein},\ and\ \citenamefont
  {Jaegermann}}]{Lohaus2018-fe2o3-polarons}%
  \BibitemOpen
  \bibfield  {author} {\bibinfo {author} {\bibfnamefont {C.}~\bibnamefont
  {Lohaus}}, \bibinfo {author} {\bibfnamefont {A.}~\bibnamefont {Klein}},\ and\
  \bibinfo {author} {\bibfnamefont {W.}~\bibnamefont {Jaegermann}},\ }\href
  {https://doi.org/10.1038/s41467-018-06838-2} {\bibfield  {journal} {\bibinfo
  {journal} {Nat. Commun.}\ }\textbf {\bibinfo {volume} {9}},\ \bibinfo {pages}
  {4309} (\bibinfo {year} {2018})}\BibitemShut {NoStop}%
\bibitem [{\citenamefont {Solyman}\ \emph {et~al.}(2022)\citenamefont
  {Solyman}, \citenamefont {Ahmed},\ and\ \citenamefont {Azab}}]{Solyman2022}%
  \BibitemOpen
  \bibfield  {author} {\bibinfo {author} {\bibfnamefont {S.}~\bibnamefont
  {Solyman}}, \bibinfo {author} {\bibfnamefont {E.~M.}\ \bibnamefont {Ahmed}},\
  and\ \bibinfo {author} {\bibfnamefont {A.~A.}\ \bibnamefont {Azab}},\ }\href
  {https://doi.org/10.1088/1402-4896/ac77c5} {\bibfield  {journal} {\bibinfo
  {journal} {Phys. Scr.}\ }\textbf {\bibinfo {volume} {97}},\ \bibinfo {pages}
  {075815} (\bibinfo {year} {2022})}\BibitemShut {NoStop}%
\bibitem [{\citenamefont {Abareshi}\ and\ \citenamefont
  {Salehi}(2022)}]{Abareshi2022}%
  \BibitemOpen
  \bibfield  {author} {\bibinfo {author} {\bibfnamefont {A.}~\bibnamefont
  {Abareshi}}\ and\ \bibinfo {author} {\bibfnamefont {N.}~\bibnamefont
  {Salehi}},\ }\href {https://doi.org/10.1007/s10854-022-09220-7} {\bibfield
  {journal} {\bibinfo  {journal} {J. Mater. Sci.: Mater. Electron.}\ }\textbf
  {\bibinfo {volume} {33}},\ \bibinfo {pages} {25153} (\bibinfo {year}
  {2022})}\BibitemShut {NoStop}%
\bibitem [{\citenamefont {{Sonia}}\ \emph {et~al.}(2023)\citenamefont
  {{Sonia}}, \citenamefont {Kumari}, \citenamefont {{Suman}}, \citenamefont
  {Chahal}, \citenamefont {Devi}, \citenamefont {Kumar}, \citenamefont {Kumar},
  \citenamefont {Kumar},\ and\ \citenamefont {Kumar}}]{Sonia2023}%
  \BibitemOpen
  \bibfield  {author} {\bibinfo {author} {\bibnamefont {{Sonia}}}, \bibinfo
  {author} {\bibfnamefont {H.}~\bibnamefont {Kumari}}, \bibinfo {author}
  {\bibnamefont {{Suman}}}, \bibinfo {author} {\bibfnamefont {S.}~\bibnamefont
  {Chahal}}, \bibinfo {author} {\bibfnamefont {S.}~\bibnamefont {Devi}},
  \bibinfo {author} {\bibfnamefont {S.}~\bibnamefont {Kumar}}, \bibinfo
  {author} {\bibfnamefont {S.}~\bibnamefont {Kumar}}, \bibinfo {author}
  {\bibfnamefont {P.}~\bibnamefont {Kumar}},\ and\ \bibinfo {author}
  {\bibfnamefont {A.}~\bibnamefont {Kumar}},\ }\href
  {https://doi.org/10.1007/s00339-022-06288-0} {\bibfield  {journal} {\bibinfo
  {journal} {Appl. Phys. A}\ }\textbf {\bibinfo {volume} {129}},\ \bibinfo
  {pages} {91} (\bibinfo {year} {2023})}\BibitemShut {NoStop}%
\bibitem [{\citenamefont {Biswas}\ \emph {et~al.}(2018)\citenamefont {Biswas},
  \citenamefont {Husek}, \citenamefont {Londo},\ and\ \citenamefont
  {Baker}}]{Biswas2018}%
  \BibitemOpen
  \bibfield  {author} {\bibinfo {author} {\bibfnamefont {S.}~\bibnamefont
  {Biswas}}, \bibinfo {author} {\bibfnamefont {J.}~\bibnamefont {Husek}},
  \bibinfo {author} {\bibfnamefont {S.}~\bibnamefont {Londo}},\ and\ \bibinfo
  {author} {\bibfnamefont {L.~R.}\ \bibnamefont {Baker}},\ }\href
  {https://doi.org/10.1021/acs.nanolett.7b04818} {\bibfield  {journal}
  {\bibinfo  {journal} {Nano Lett.}\ }\textbf {\bibinfo {volume} {18}},\
  \bibinfo {pages} {1228} (\bibinfo {year} {2018})}\BibitemShut {NoStop}%
\bibitem [{\citenamefont {Li}\ \emph {et~al.}(2010)\citenamefont {Li},
  \citenamefont {Dai}, \citenamefont {Zhou}, \citenamefont {Zhang},
  \citenamefont {Wan}, \citenamefont {Fu}, \citenamefont {Zhang}, \citenamefont
  {Liu}, \citenamefont {Cao}, \citenamefont {Pan}, \citenamefont {Zhang},\ and\
  \citenamefont {Zou}}]{Li2010}%
  \BibitemOpen
  \bibfield  {author} {\bibinfo {author} {\bibfnamefont {Y.}~\bibnamefont
  {Li}}, \bibinfo {author} {\bibfnamefont {G.}~\bibnamefont {Dai}}, \bibinfo
  {author} {\bibfnamefont {C.}~\bibnamefont {Zhou}}, \bibinfo {author}
  {\bibfnamefont {Q.}~\bibnamefont {Zhang}}, \bibinfo {author} {\bibfnamefont
  {Q.}~\bibnamefont {Wan}}, \bibinfo {author} {\bibfnamefont {L.}~\bibnamefont
  {Fu}}, \bibinfo {author} {\bibfnamefont {J.}~\bibnamefont {Zhang}}, \bibinfo
  {author} {\bibfnamefont {R.}~\bibnamefont {Liu}}, \bibinfo {author}
  {\bibfnamefont {C.}~\bibnamefont {Cao}}, \bibinfo {author} {\bibfnamefont
  {A.}~\bibnamefont {Pan}}, \bibinfo {author} {\bibfnamefont {Y.}~\bibnamefont
  {Zhang}},\ and\ \bibinfo {author} {\bibfnamefont {B.}~\bibnamefont {Zou}},\
  }\href {https://doi.org/10.1007/s12274-010-1036-y} {\bibfield  {journal}
  {\bibinfo  {journal} {Nano Res.}\ }\textbf {\bibinfo {volume} {3}},\ \bibinfo
  {pages} {326} (\bibinfo {year} {2010})}\BibitemShut {NoStop}%
\bibitem [{\citenamefont {Thi Lan~Huong}\ \emph {et~al.}(2023)\citenamefont
  {Thi Lan~Huong}, \citenamefont {Van~Quang}, \citenamefont {Thi~Huyen},
  \citenamefont {Thu~Huong}, \citenamefont {Anh~Tuan}, \citenamefont
  {Trung~Tran}, \citenamefont {Vinh~Tran}, \citenamefont {Ngoc~Bach},
  \citenamefont {Tu},\ and\ \citenamefont {Dao}}]{ThiLanHuong2023}%
  \BibitemOpen
  \bibfield  {author} {\bibinfo {author} {\bibfnamefont {P.}~\bibnamefont {Thi
  Lan~Huong}}, \bibinfo {author} {\bibfnamefont {N.}~\bibnamefont {Van~Quang}},
  \bibinfo {author} {\bibfnamefont {N.}~\bibnamefont {Thi~Huyen}}, \bibinfo
  {author} {\bibfnamefont {H.}~\bibnamefont {Thu~Huong}}, \bibinfo {author}
  {\bibfnamefont {D.}~\bibnamefont {Anh~Tuan}}, \bibinfo {author}
  {\bibfnamefont {M.}~\bibnamefont {Trung~Tran}}, \bibinfo {author}
  {\bibfnamefont {Q.}~\bibnamefont {Vinh~Tran}}, \bibinfo {author}
  {\bibfnamefont {T.}~\bibnamefont {Ngoc~Bach}}, \bibinfo {author}
  {\bibfnamefont {N.}~\bibnamefont {Tu}},\ and\ \bibinfo {author}
  {\bibfnamefont {V.-D.}\ \bibnamefont {Dao}},\ }\href
  {https://www.sciencedirect.com/science/article/pii/S0038092X22008829}
  {\bibfield  {journal} {\bibinfo  {journal} {J. Sol. Energy}\ }\textbf
  {\bibinfo {volume} {249}},\ \bibinfo {pages} {712} (\bibinfo {year}
  {2023})}\BibitemShut {NoStop}%
\bibitem [{\citenamefont {Mahdy}\ \emph {et~al.}(2022)\citenamefont {Mahdy},
  \citenamefont {Azab}, \citenamefont {El~Zawawi},\ and\ \citenamefont
  {Turky}}]{Mahdy2022-ZnO-CFO}%
  \BibitemOpen
  \bibfield  {author} {\bibinfo {author} {\bibfnamefont {M.~A.}\ \bibnamefont
  {Mahdy}}, \bibinfo {author} {\bibfnamefont {A.~A.}\ \bibnamefont {Azab}},
  \bibinfo {author} {\bibfnamefont {I.~K.}\ \bibnamefont {El~Zawawi}},\ and\
  \bibinfo {author} {\bibfnamefont {G.}~\bibnamefont {Turky}},\ }\href
  {https://doi.org/10.1088/1402-4896/aca5bc} {\bibfield  {journal} {\bibinfo
  {journal} {Phys. Scr.}\ }\textbf {\bibinfo {volume} {98}},\ \bibinfo {pages}
  {015806} (\bibinfo {year} {2022})}\BibitemShut {NoStop}%
\end{thebibliography}

\providecommand{\noopsort}[1]{}\providecommand{\singleletter}[1]{#1}%

\end{document}